\documentstyle[aps,prd]{revtex}

\input epsf

\begin{document}

\draft

\preprint{LAEFF-96/07, gr-qc/9604019}

\title{Understanding critical collapse of a scalar field}

\author{Carsten Gundlach}

\address{LAEFF-INTA (Laboratorio de Astrof\'\i sica Espacial y F\'\i sica
Fundamental -- Instituto Nacional de Tecnolog\'\i a Aerospacial), 
PO~Box~50727, 28080~Madrid, Spain}

\date{8 April, 1996}

\maketitle


\begin{abstract}

I construct a spherically symmetric solution for a massless real
scalar field minimally coupled to general relativity which is
discretely self-similar (DSS) and regular. This solution coincides
with the intermediate attractor found by Choptuik in critical
gravitational collapse. The echoing period is $\Delta = 3.4453 \pm
0.0005$.  The solution is continued to the future self-similarity
horizon, which is also the future light cone of a naked
singularity. The scalar field and metric are $C^1$ but not $C^2$ at
this Cauchy horizon. The curvature is finite nevertheless, and the
horizon carries regular null data. These are very nearly flat. The
solution has exactly one growing perturbation mode, thus confirming
the standard explanation for universality. The growth of this mode
corresponds to a critical exponent of $\gamma = 0.374 \pm 0.001$, in
agreement with the best experimental value. I predict that in critical
collapse dominated by a DSS critical solution, the scaling of the
black hole mass shows a periodic wiggle, which like $\gamma$ is
universal. My results carry over to the free complex scalar
field. Connections with previous investigations of self-similar scalar
field solutions are discussed, as well as an interpretation of
$\Delta$ and $\gamma$ as anomalous dimensions.

\end{abstract}

\pacs{04.25.Dm, 04.20.Dw, 04.40.Nr, 04.70.Bw, 05.70.Jk}


\section{Introduction}


\subsection{Critical phenomena in gravitational collapse}


In an astrophysical context, gravitational collapse normally starts
from a star. This means that the initial data are almost stationary,
and that they have a characteristic scale which is provided by the
matter. Therefore astrophysical black holes have a minimum mass,
namely the Chandrasekhar mass. Abandoning the restriction to almost
stationary initial data, or alternatively to realistic matter, one
should be able to make arbitrarily small black holes. One may then ask
what happens if one tries to make a black hole of infinitesimal mass
by fine-tuning the initial data.

The investigation is simplified by choosing a matter model that does not
admit stable stationary solutions. Then, for {\it any} initial data, there
are only two possible outcomes: formation of a black hole, or dispersion
leaving behind flat spacetime.  The first systematic numerical examination
of the limit between the two (the ``critical surface'' in phase space) was
carried out by Choptuik \cite{Chop} for a massless minimally coupled real
scalar field in spherical symmetry. He evolved members of various
one-parameter families of initial data each of which comprised both
collapsing and dispersing data, and searched for the critical parameter
value by bisection.  For all families he investigated he was able to make
arbitrarily small black holes by tuning the parameter $p$ of the data:
there was no evidence for a ``mass gap''. Instead he found two unexpected
new phenomena.

For marginal data, both supercritical and subcritical, the time evolution
approaches a certain universal solution which is the same for all the
one-parameter families of data. This solution is an ``intermediate
attractor'' in the sense that the time evolution first converges onto it,
but then diverges from it eventually, to either form a black hole
or to disperse. This universal solution (also called the ``critical
solution'') has a curious symmetry: it is periodic in the logarithm of
spacetime scale, with a period of $\Delta \simeq 3.44$. This is also referred
to as ``echoing'', or discrete self-similarity (DSS).

Moreover, for marginally {\it super}critical data, the final black hole mass
scales as $M \sim (p-p_*)^\gamma$, where $p$ is the parameter of the family
of initial data, and $p_*$ its critical value. The ``critical exponent''
$\gamma$ has the value $\simeq 0.37$ for the scalar field, and
like the critical solution it is universal in the sense that it is the same
for all one-parameter families of data.

Both phenomena were then also found in the axisymmetric collapse of
pure gravity \cite{AbrEv}, indicating that they are an artifact of
neither the choice of matter nor of spherical symmetry. There,
$\Delta$ was found to be $\simeq 0.6$, and the critical exponent
$\gamma$ to be $\simeq 0.36$.  For a perfect fluid with equation of
state $p=\rho/3$ in spherical symmetry \cite{EvCol}, the universal
attractor has a different symmetry: it is not discretely, but
continuously self-similar (CSS). $\gamma$ is found to be $\simeq 0.36$
once more.

Choptuik's results have been duplicated, to varying precision, in
\cite{GPP2,Garf,HamStew}.

Subsequently, the matter models were generalized. For a fluid with
$p=k\rho$ in spherical symmetry, now with arbitrary constant $k$,
$\gamma$ was found to be strongly $k$-dependent \cite{Maison}
\footnote{The $\gamma$ for $k\ne 1/3$ have been calculated only in linear
perturbation analysis, but I strongly expect them to be confirmed by
collapse calculations.}.  The real scalar field model was generalized to a
one-parameter family of two-component non-linear sigma models
\cite{HE3}. This family includes the cases of a free complex scalar field
\cite{HE1,HE2}, a real scalar field coupled to Brans-Dicke gravity
\cite{ChopLieb} and, as a special case of the latter, a string-inspired
axion-dilaton model \cite{EHH}.  For the axion-dilaton model, $\gamma
\simeq 0.264$ is found in collapse simulations\cite{HHS}. From these new
examples it is clear that $\gamma$ is not universal with respect to
different kinds of matter, but only with respect to the initial data for
any one matter model.


\subsection{The emerging picture}


Let us now examine some general features of the ``critical solutions''
which appear as intermediate attractors in collapse, and which seem to
describe the limiting case of the formation of a zero mass black
hole. First of all, they must be scale-invariant in some way, and in
fact will show homotheticity, or ``self-similarity of the first kind''
\cite{CarHen}. Because of self-similarity, they must have a curvature
singularity, but they should not have an event horizon. The absence of
a horizon means that the collapsing matter is visible in the solution.

The unique endpoint of gravitational collapse is given by the
Kerr-Newman family of solutions because, roughly speaking, they only
admit damped perturbations: they are attractors in phase space. If the
critical collapse solution were also attractors, we would see many
naked singularities in nature, and this is clearly not the case. In
fact they are attractors of co-dimension one, and we shall see that
this gives rise to an analogue of the ``no hair'' theorem:
universality.  The Kerr-Newman and critical collapse solutions are
briefly contrasted in Table \ref{table1}.

\begin{figure}
\epsfysize=16cm
\centerline{\epsffile{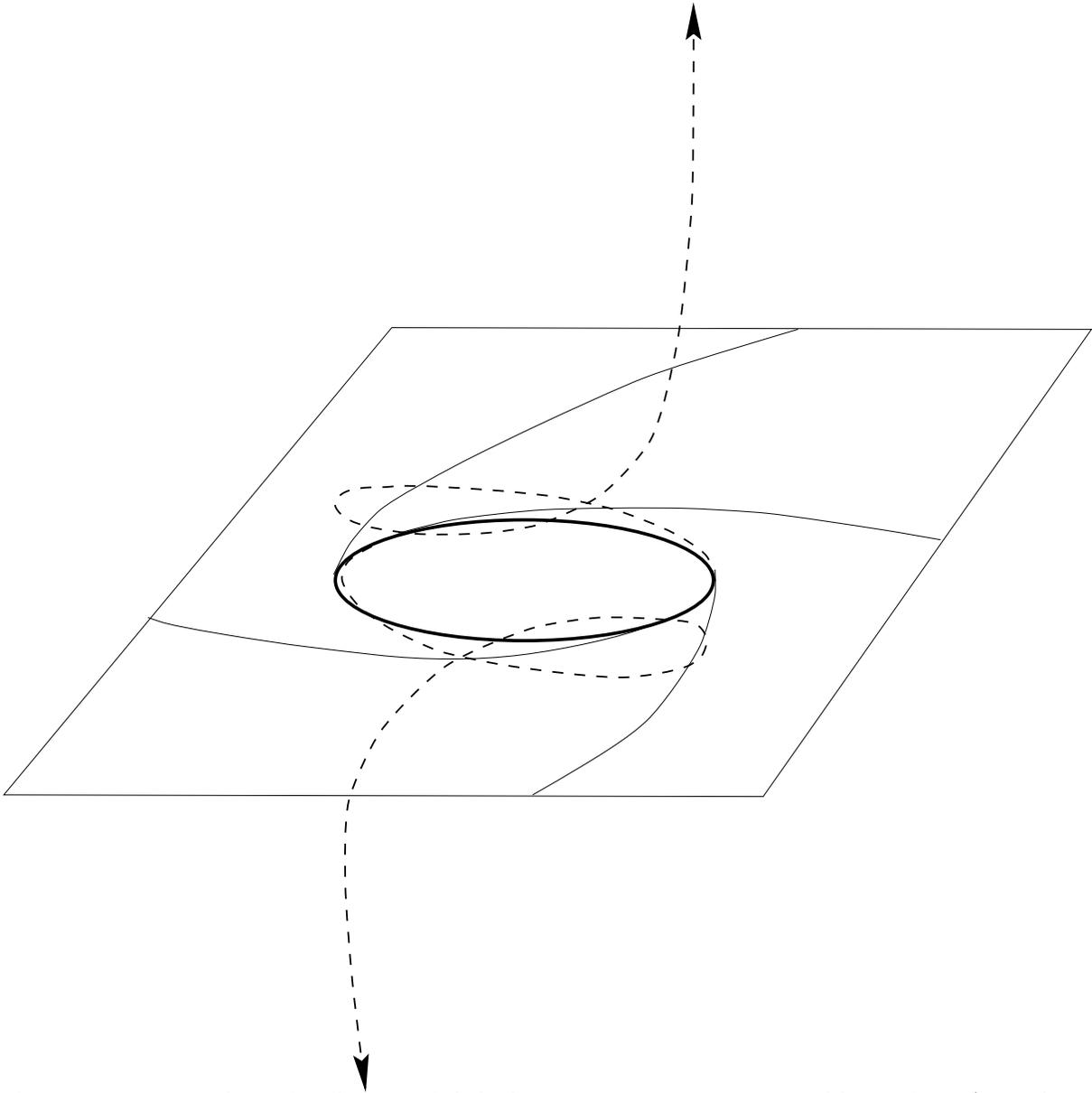}}
\caption{The phase space picture for discrete self-similarity. The plane
represents the critical surface. (In reality this is a hypersurface of
co-dimension one in an infinite-dimensional space.) The circle (fat
unbroken line) is the limit cycle representing the critical solution. The
thin unbroken curves are spacetimes attracted to it. The dashed curves are
spacetimes repelled from it. There are two families of such curves, labeled
by one periodic parameter, one forming a black hole, the other dispersing
to infinity. Only one member of each family is shown.}
\label{phasespace}
\end{figure}

In a schematic picture of phase space \cite{Koike}, the critical
solution remains within the critical surface. The observed
universality suggests that it is in fact an attractor within the
critical surface, and in consequence an attractor of co-dimension one
in phase space. This attractor could either be a fixed point,
corresponding to CSS, or a limit cycle, corresponding to DSS. Fig.\
\ref{phasespace} illustrates this. 

Nearly critical Cauchy data are situated close to the critical
surface, but may be far from the critical point. Under time evolution
they are attracted towards the critical point. While they approach it,
their ``distance'' from the critical surface increases exponentially
but remains small until they are close to the critical point because,
by the assumption of near-criticality, it is initially very small. Near
the critical point, the exponential increase takes over, and the phase
space trajectories are repelled from the critical surface, all into
the same direction. This constitutes the mechanism of universality
with respect to initial data. The formula for the black hole mass
follows essentially by dimensional analysis, and the critical exponent
$\gamma$ can be read off from the linear perturbations of the critical
solution \cite{Koike}. As we shall see in section \ref{subsec:scaling}, the
periodicity of the critical solution in the DSS case gives rise to a
periodic wiggle in the scaling law.

What confuses the naive phase space picture is gauge-invariance: the
same spacetime corresponds to very different trajectories in
superspace, depending on how it is sliced. The naive picture must
therefore be used with care pending its formulation in geometric
terms.  

CSS solutions and their linear perturbations have already been
calculated for various matter models.  For the $p=\rho/3$ model \cite
{EvCol,Koike} and for the axion-dilaton model \cite{EHH,HHS} they
agree with the experimentally determined critical solution and give
the correct critical exponent. For the complex scalar field a CSS
solution has been found \cite{HE1}, but it is only an attractor of
co-dimension three \cite{HE2}. For the $p=k\rho$ model with $k\ne
1/3$, $\gamma$ has been calculated formally on on the basis of the
perturbations of a CSS solution\cite{Maison}, but it is not known for
what range of $k$ this CSS solution is really the critical
solution. (By continuity, it must be for some neighborhood of
$k=1/3$.) Furthermore, no CSS solution exists for $k \gtrsim 0.888$
\cite{Maison}, and one would expect that the critical solution becomes
DSS at either that, or a smaller, value of $k$.  The CSS solution and
its perturbations have been calculated also for the family of
nonlinear sigma models \cite{HE3}. The number of unstable modes of the
CSS solution changes from one to three at some value of the
parameter. This probably indicates the changeover from CSS to DSS in
the critical solution. Why some critical solutions are CSS and others
DSS is not yet understood, however.

The present paper gives the first calculation of a DSS critical solution,
together with its maximal extension and its linear perturbations. This is
technically much more difficult than CSS, but DSS is the most generic case
of self-similarity, and the mathematical and numerical methods developed
here will be useful in other applications. The critical mass scaling
generalizes to contain a universal periodic wiggle.

The plan of the paper is as follows. In section \ref{sec:critsol}, I
define DSS and construct the DSS solution of the real scalar field
model in the past light cone of the singularity as a nonlinear
eigenvalue problem. It agrees with the critical solution found by
Choptuik \cite{Chop}. This section, together with appendices
\ref{appA}, \ref{appC}, \ref{appD} and \ref{appF}, is an expansion of
the Letter \cite{PRL}. Sections \ref{sec:maxext} and \ref{sec:pert}
both build on section \ref{sec:critsol}, but are independent of each
other.

In section \ref{sec:maxext}, I extend the critical solution up to the
future light cone of the naked singularity. I find that this Cauchy
horizon is regular, and that it is in fact nearly flat.  In section
\ref{sec:pert}, I calculate the linear perturbations, and generalize
the calculation of the critical exponent to the DSS case. My value for
$\gamma$ agrees with the experimental one, but I also predict the
existence of a (small) universal wiggle overlaid on the power-law
scaling, which has not been observed so far. I show that the Choptuik
solution is the critical solution not only for the real, but also the
free complex scalar field.  In section \ref{sec:concl}, I put my
results into the context of results for other collapsing systems on
one hand, and of the study of (continuously) self-similar scalar field
models on the other. I then discuss the next steps to be taken, and
the possible connection with critical phenomena in statistical
mechanics and quantum field theory. Various details are given in the
appendices.


\section{The critical solution}
\label{sec:critsol}


\subsection{Field equations}


In this section I construct an isolated solution of general relativity
minimally coupled to a massless real scalar field with the following
properties: 1) spherical symmetry, 2) discrete self-similarity (to be
defined below), 3) analytic at the center of spherical symmetry, 4)
analytic at the past self-similarity horizon, 5) the scalar field is
bounded. (It is possible that there is no solution obeying 1) to 4) that
does not also obey 5), but I have not shown this.)

The Einstein equations we consider here are
\begin{equation} 
G_{ab}=8\pi G\, \left(\phi_{,a}\phi_{,b}-{1\over2}g_{ab}
\phi_{,c}\phi^{,c}\right),
\end{equation}
in spherical symmetry.  The matter equation $\phi_{,c}^{\ \ ;c}=0$
follows from the Einstein equations as the contracted Bianchi identity.
The spacetime metric is
\begin{equation} 
\label{metric}
ds^2\equiv-\alpha(r,t)^2\,dt^2+a(r,t)^2\,dr^2+r^2\,
(d\theta^2+\sin^2\theta\,d\varphi^2).
\end{equation}
This form of the metric is invariant under
transformations $t \to \tilde t(t)$, $\alpha \to \tilde \alpha$, such
that $\alpha \, dt = \tilde \alpha \, d\tilde t$. In
order to write the matter equations in first order 
form, we introduce the auxiliary matter fields
\begin{equation}
X(r,t)\equiv\sqrt{2\pi G}\ {r\over a}\phi_{,r},\qquad
Y(r,t)\equiv\sqrt{2\pi G}\ {r\over \alpha}\phi_{,t}.
\end{equation}
The combinations $X_\pm = X \pm Y$ of these fields propagate along
characteristics.  The radial null geodesics, which are also the matter
characteristics, are characterized by the quantity $g\equiv a/\alpha$
\footnote{I have made one change of notation. In \cite{PRL} I
defined $g\equiv \exp(\xi_0) a/\alpha$, while $g \equiv
a/\alpha$ here. This is for greater convenience in the remainder of the
paper.}.  
The scalar wave equation in $X_\pm$ is then
\begin{equation}
r \left( X_{\pm,r} \mp g X_{\pm,t} \right) = \left[
\pm r {a_{,t} \over \alpha} - r {\alpha_{,r} \over \alpha}
\right] X_\pm - X_\mp.
\end{equation}
In the following we use $X_+$, $X_-$, $g$ and $a$ as our basic
variables. A complete set of Einstein equations in these variables is
\begin{eqnarray}
\label{first}
r g_{,r} && = (1 - a^2) g, \\
r a_{,r} && = {1 \over 2} a \left[(1 - a^2) + a^2 (X_+^2 + X_-^2)\right]
\equiv C_1, \\
g r a_{,t} && = {1 \over 2} a^3 (X_+^2 - X_-^2) \equiv C_2, 
\end{eqnarray}
and the matter equations become
\begin{equation}
\label{last}
r \left( X_{\pm,r} \mp g X_{\pm,t} \right) = \left[{1 \over 2} (1 - a^2) - a^2
X_\mp^2\right] X_\pm - X_\mp \equiv C_\pm
\end{equation}
when we eliminate the metric derivatives with the help of the Einstein
equations. The five first order equations (\ref{first}-\ref{last}) are our
field equations.  I have defined the expressions $C_\pm$, $C_1$ and $C_2$
for later convenience. The absence of $g_{,t}$ in the equations reflects
the fact that $\alpha$, and hence $g$, contains a gauge degree of freedom
not determined by the Cauchy data.

The two scalars made from the Ricci curvature, using the Einstein
equations, are
\begin{equation}
R = 4 r^{-2} X_+ X_-,\qquad R_{ab} R^{ab} = R^2.
\end{equation}
The Riemann tensor will be considered in appendix \ref{appG}.


\subsection{Discrete self-similarity}


The concept of (continuous) self-similarity (CSS) (or homotheticity)
has been defined in a relativistic context \cite{CahTaub,CarHen} as
the presence of a vector field $\chi$ such that
\begin{equation}
{\cal L}_\chi g_{ab} = 2 g_{ab},
\end{equation}
where ${\cal L}_\chi$ denotes the Lie derivative. I now introduce the concept
of discrete self-similarity (DSS). In this symmetry there exist a
diffeomorphism $\phi$ and a real constant $\Delta$ such that, for any
integer $n$,
\begin{equation} 
\label{discrete}
\left(\phi_*\right)^n g_{ab} = e^{2n\Delta} g_{ab},
\end{equation}
where $\phi_*$ is the pull-back of $\phi$.

To see what DSS looks like in coordinate terms, we introduce coordinates
$(\sigma,x^\alpha)$, such that if a point $p$ has coordinates
$(\sigma,x^\alpha)$, its image $\phi(p)$ has coordinates
$(\sigma+\Delta,x^\alpha)$. One can verify that DSS in these
coordinates is equivalent to
\begin{equation}
\label{rescaled}
g_{\mu\nu}(\sigma,x^\alpha) = e^{2\sigma} \tilde
g_{\mu\nu}(\sigma,x^\alpha),
\quad \hbox{where} \quad \tilde
g_{\mu\nu}(\sigma,x^\alpha) = \tilde
g_{\mu\nu}(\sigma+\Delta,x^\alpha)
\end{equation}
In other words, the DSS acts as a discrete isomorphism on the rescaled
metric $\tilde g_{\mu\nu}$. $\sigma$ is intuitively speaking the logarithm
of spacetime scale. 

One can formally construct such a coordinate system in the following
way: Fix a hypersurface $\Sigma$ such that its image $\Sigma'$ under
$\phi$ does not intersect $\Sigma$. Introduce coordinates $x^\alpha$
in $\Sigma$, and copy them to $\Sigma'$ with $\phi$. Introduce
coordinates $(\sigma,x^\alpha)$ in the region between $\Sigma$ and
$\Sigma'$ such that $\sigma$ has range $[0,\Delta]$, their restriction
to $\Sigma$ is $(0,x^\alpha)$ and their restriction to $\Sigma'$ is
$(\Delta,x^\alpha)$. Finally, copy these coordinates to the entire
spacetime, such that if $p$ has coordinates $(\sigma,x^\alpha)$, its
$n$-th image $\phi^n(p)$ is assigned coordinates
$(\sigma+n\Delta,x^\alpha)$. Clearly there is enormous freedom in
defining such coordinates.

In order to clarify the connection between CSS and DSS, one may define
a vector field $\chi\equiv \partial / \partial \sigma$, although there
is no unique $\chi$ associated with a given $\phi$. The discrete
diffeomorphism $\phi$ is then realized as the Lie dragging along
$\chi$ by a distance $\Delta$. Clearly, CSS corresponds to DSS for
infinitesimally small $\Delta$, and hence for all $\Delta$, and is in
this sense a degenerate case of DSS. In this limit, $\chi$ becomes
unique. In the following I speak of DSS only in the absence of CSS.

In order to see what form DSS takes in spherical symmetry in the particular
coordinates defined in (\ref{metric}), we make a coordinate transformation
$t\equiv e^\sigma
T(\sigma,z)$ and $r\equiv e^\sigma
R(\sigma,z)$, where $T$ and $R$ are periodic in $\sigma$ with period
$\Delta$. Here $\sigma$ is the same as in the general construction, and
$x^\alpha=(z,\theta,\varphi)$. 
In the new coordinates the
metric (\ref{metric}) takes the form
\begin{equation}
\label{metric2}
ds^2 = e^{2\sigma} \left\{-\alpha^2 
\left[(T+T_{,\sigma})\,d\sigma + T_{,z}\, dz\right]^2 + a^2
\left[(R+R_{,\sigma})\,d\sigma + R_{,z}\, dz\right]^2 +
R^2\left(d\theta^2 + \sin^2\theta \, d\varphi^2\right) \right\}
\end{equation}
This is of the form (\ref{rescaled}) if and only if $\alpha(z,\sigma)$ and
$a(z,\sigma)$ are also periodic in $\sigma$ with period $\Delta$. In terms
of $t$ and $r$ this periodicity corresponds to
\begin{equation}
\label{echo}
a(r, t) = a(e^{n\Delta} r, e^{n\Delta} t), \qquad
\alpha(r, t) = \alpha(e^{n\Delta} r, e^{n\Delta} t).
\end{equation}
This is not yet the most general way to impose DSS in
(\ref{metric}). We obtain that by also admitting the transformations
$t\to \bar t(t)$, with $\alpha \to \bar \alpha = dt/d\bar t\
\alpha$. $\bar \alpha$ need no longer be periodic, but it must be
related in this way to some $\alpha$ that is.


\subsection{Formulation as an eigenvalue problem}


We now introduce specific coordinates of the kind just discussed. The
following choice will be
sufficiently general for our purpose:
\begin{equation}
\label{tauzeta}
\tau \equiv \ln \left({t\over r_0}\right), \qquad \zeta \equiv \ln\left({r\over
t}\right) - \xi_0(\tau).
\end{equation}
Here $r_0$ is an arbitrary fixed scale, and $\xi_0(\tau)$ is a
periodic function with period $\Delta$\footnote{As in \cite{PRL}, I
assume $t>0$. In a collapse context, where the spacetime region we are
about to calculate is to the past of the singularity, $t$ then
decreases to the future, but this is purely a matter of
convention. The convention can be reversed in the results simply by
changing the sign of $Y$.}. Both are to be determined later.  In the
new coordinates, the matter and Einstein equations are
\begin{eqnarray}
\label{matter}
X_{\pm,\zeta} && = {C_\pm \pm e^{\zeta + \xi_0} g X_{\pm,\tau}
\over 1 \pm (1+\xi_0') e^{\zeta + \xi_0} g} \\
\label{1}
g_{,\zeta} && = (1-a^2) g \\
\label{2}
a_{,\zeta} && = C_1 \\
\label{constraint}
a_{,\tau} && = e^{-(\zeta + \xi_0)} g^{-1} C_2 + (1+\xi_0') C_1 
\end{eqnarray}
Here $\xi_0' = d\xi_0(\tau) / d\tau$, and in this paper a prime always
denotes the derivative of a function of one variable with respect to
its formal argument. The equations are invariant under a translation
in $\tau$, corresponding to a change in the arbitrary scale $r_0$.

In order to impose discrete homotheticity, we demand the periodicity
(\ref{echo}).  We also impose the regularity condition $a = 1$ at
$r=0$ and the gauge condition $\alpha = 1$ at $r=0$. (Both are
compatible with the periodicity.) In our choice of dependent and
independent variables we therefore impose boundary conditions
\begin{equation} 
\label{periodic}
a(\zeta, \tau + n \Delta) = a(\zeta, \tau), \qquad
g(\zeta, \tau + n \Delta) = \alpha(\zeta, \tau),
\end{equation}
and
\begin{equation}
a(\zeta = -\infty, \tau) = 1, \qquad g(\zeta = -\infty, \tau) = 1.
\end{equation}
Note that we continue to describe the metric with the variables $a$
and $\alpha$, but that they are not the coefficients of the metric
associated with coordinates $(\tau,\zeta)$.

From the Einstein equations it follows that the periodicity condition must
hold also for $X_+$ and $X_-$. From the equations defining $X$ and $Y$, we
obtain
\begin{equation}
\label{phi}
\phi_{,\tau} = (2 \pi G)^{-1/2} a \left[ (1 + \xi_0') X + g^{-1} e^{-(\zeta + \xi_0)} Y
\right], \qquad \phi_{,\zeta} = (2 \pi G)^{-1/2} a X.
\end{equation}
Because the right-hand sides of both equations are periodic in $\tau$, the
scalar field $\phi$ itself is of the form 
\begin{equation}
\phi(\zeta, \tau + \Delta) = \hbox{(periodic in $\tau$)} + \kappa \tau,
\end{equation}
where $\kappa$ is a constant. $\kappa$ is not an
independent parameter, but is determined through
the first of equations (\ref{phi}). $\phi$ is bounded and
periodic if and only if $\kappa=0$.

As we have a $1+1$ dimensional hyperbolic problem, we can interchange
space and time. In this view, near $r=0$, equations (\ref{matter}),
(\ref{1}) and (\ref{2}) form a first-order system of evolution
equations for $a$, $g$, $X_+$ and $X_-$ with ``time'' coordinate
$\zeta$ and (periodic) ``space'' coordinate $\tau$. The data are
subject to one constraint (\ref{constraint}), which is propagated by
the evolution equations. As $\zeta\to-\infty$, we impose boundary
conditions corresponding to a regular center $r=0$.

The equations become singular when the denominator of (\ref{matter})
vanishes. The treatment of this singular surface, both analytical and
numerical, is simplified if we use the coordinate freedom incorporated in
the choice of $\xi_0(\tau)$ to make this happen ``for all $\tau$ at once'',
namely on the line $\zeta = 0$. We therefore impose the coordinate
condition
\begin{equation}
\label{coordcond}
\left[ 1 - (1+\xi_0') e^{\xi_0} g \right]_{\zeta=0}  = 0.
\end{equation}
A necessary condition for regularity is now that the denominator also
vanishes at $\zeta = 0$, or
\begin{equation}
\label{regularity}
\left[C_- - e^{\xi_0} g X_{-,\tau}\right]_{\zeta = 0} = 0.
\end{equation}
We may look at this from a different angle: The coordinate condition means
in fact that $\zeta=0$ is null, and hence a characteristic of the
equations. Such a characteristic which is mapped onto itself by the
self-similarity map $\phi$ is called a ``self-similarity
horizon'' (SSH), or a ``sonic line'' \cite{Selfsim}. For our equations,
with periodic boundary conditions in the ``space'' coordinate $\tau$, it
constitutes a Cauchy horizon. On every other $\zeta={\rm const.}$ surface,
$X_+$ and $X_-$ would be free Cauchy data, but on $\zeta=0$ they are
overdetermined, as one requires only characteristic data. This gives again
rise to the condition (\ref{regularity}).

We still have to see what kind of regularity this condition enforces,
and if it is sufficient as well as necessary.  To investigate the
possible behavior of the solution at the past SSH $\zeta=0$, we assume
for the moment that $a$, $g$ and $X_+$ are at least $C^0$ there, and
expand the equation for $X_-$ to leading order in $\zeta$. The
resulting approximate equation, using the coordinate condition
(\ref{coordcond}) and equation (\ref{1}), is
\begin{equation}
\label{approx}
X_{-,\zeta} = {A(\tau) X_- - X_{-,\tau} + C(\tau) \over \zeta B(\tau)},
\end{equation}
where the coefficients
\begin{equation}
A = (1+\xi_0') \left[{1\over2} (1-a_0^2) - a_0^2 X_{+0}^2\right],
\qquad B = - (1+\xi_0') (1-a_0^2)
\qquad C = - (1+\xi_0') X_{+0}
\end{equation}
are evaluated at $\zeta=0$. This approximate equation admits an exact
general solution, namely
\begin{equation}
X_- = X_-^{{\rm inhom}}(\tau) + X_-^{{\rm hom}}(\zeta,\tau).
\end{equation}
The particular inhomogeneous solution $X_-^{{\rm inhom}}$ is defined
as the unique solution of
\begin{equation}
A X_-^{{\rm inhom}} - X_{,\tau}^{{\rm inhom}} + C = 0
\end{equation}
with periodic boundary conditions. This solution exists and is unique,
unless the average value of $A$ vanishes. The general homogeneous
solution $X_-^{{\rm hom}}$ is of the form
\begin{equation}
X_-^{{\rm hom}} = \zeta^{\hat A_0\over \hat B_0} \ e^{I_0 A - {\hat A_0\over
\hat B_0} I_0 B} \ F\left[\tau + {I_0 B - \ln |\zeta| \over \hat B_0}\right].
\end{equation}
The notation here is that of appendix \ref{appC}: $\hat A_0$ is the
constant part, or average value, of the periodic function $A$.  $I_0 A$
denotes the principal function of $A-\hat A_0$, that is of $A$ minus its
average, with the integration constant defined so that $I_0 A$ itself has
zero average. $F$ is a periodic function of one variable with period
$\Delta$. It is the free parameter of the solution. We do not need to
determine it here.  What is important is that the solution is analytic if
and only if $F$ vanishes identically. If $F$ does not vanish, there are two
possibilities. If $\hat A_0/ \hat B_0 < 0$, the solution will blow up at
$\zeta=0$. If $\hat A_0/ \hat B_0 > 0$, the solution will be $C_0$ but not
$C_1$ there. It is easy to see that both $\hat A_0$ and $\hat B_0$ are
negative in our case, so that the solution either blows up or is analytic.
So with the one condition (\ref{regularity}) we automatically enforce
analyticity.

We shall impose one extra symmetry on our ansatz.  The results of
numerical collapse simulations \cite{Chop} indicate that $\phi$ itself is
periodic.  Moreover, even if one adds a potential to the scalar field
action, the attractor is found to be the same as for the massless field
\cite{Chop2}. This requires that $\phi$ remains bounded in the critical
solution, because only then can a polynomial of $\phi$ be neglected with
respect to the space and time derivatives of $\phi$, which are unbounded
because of the echoing on an exponentially decreasing spacetime scale.  For
this reason we look for a solution with $\kappa=0$.
Because all frequencies are coupled through the nonlinearity of the field
equations, $\kappa$ vanishes if and only if the even frequencies of $X_\pm$
and the odd frequencies of $a$ and $g$ vanish for all $\zeta$. Therefore we
now impose in our ansatz that $a$ and $g$, as well as $\xi_0$, are composed
only of even frequency terms in $\tau$, and $X_\pm$ only of odd frequency
terms. This reflection-type symmetry is compatible with the field
equations.

We have now completed the formulation of a hyperbolic boundary value
problem.  Its Cauchy data are values for the four fields $a$, $g$, $X_+$
and $X_-$, up to one constraint, and up to a translation invariance in
$\tau$, plus the unknown value of $\Delta$ and the unknown function
$\xi_0(\tau)$. The count is therefore $4\infty - \infty - 1 + 1 + \infty =
4 \infty$. (Here $\infty$ stands for the countably infinite number of
degrees of freedom of a periodic function of one variable.)  These free
data are balanced by two boundary conditions at $\zeta=-\infty$ and two at
$\zeta=0$, or $4\infty$ degrees of freedom again. One would therefore
expect this boundary value problem to have at most a discrete set of
solutions.

Numerically I have constructed one such solution. Locally it is unique. To
solve the boundary value problem numerically, I have expanded all periodic
fields in Fourier components of $\tau$, truncating the expansion at some
relatively small number of components. This takes advantage of the fact
that the solution is smooth, so that the Fourier series converges rapidly.
The field $a$ is not evolved in $\zeta$, but reconstructed at each step
from the constraint (\ref{constraint}). Details are given in the
appendices. I find that $\Delta = 3.4453\pm 0.0005$. In a previous
publication \cite{PRL}, I had given a value of $\Delta = 3.4439 \pm
0.0004$. The difference, corresponding to $3.5$ standard errors, is due to
a change in numerical details of the algorithm. These changes correct what
I now regard as an inconsistency in my original Pseudo-Fourier method, and
are therefore ``systematic error''.  The quoted error is in both cases
estimated discretisation error, which is discussed in appendix
\ref{appF}. The present Fourier methods are outlined in appendix
\ref{appC}.

I have shown in \cite{PRL} that the DSS solution I have constructed
agrees with the intermediate attractor observed by Choptuik
\cite{Chop} to the numerical precision of the latter. The error in the
DSS solution I have corrected here is small enough not to affect this
agreement.


\section{Maximal extension of the critical spacetime}
\label{sec:maxext}


\subsection{From the past self-similarity horizon to $t=0$}


The coordinates $(\zeta,\tau)$ become singular at $t=0$ ($\zeta =
\infty$). Clearly it will be necessary to replace $\tau$ with $\rho$
as the periodic coordinate $\sigma$ to regularize the equations there.
Before we do this, it is useful to examine the asymptotic behavior of
the solution in the old variables as $\zeta \to \infty$, or $t\to
0$. Neglecting terms of order $\exp (-\zeta)$, the field equations in
this limit reduce to
\begin{eqnarray}
(1+\xi_0') X_{\pm,\zeta} && = X_{\pm,\tau}, \\
(1+\xi_0') a_{,\zeta} && = a_{,\tau}, \\
a_{,\zeta} && = C_1, \\
g_{,\zeta} && = (1-a^2) g, 
\end{eqnarray}
with the general solution
\begin{eqnarray}
X_\pm && = X_{\pm0}(\rho), \\
a && = a_{0}(\rho), \\
\label{ginf}
g && = g_0(\tau) \, e^{\sigma(\rho) + \nu \zeta},
\end{eqnarray}
where
\begin{equation}
\label{rho}
\rho \equiv \zeta + \tau + \xi_0(\tau) \equiv \ln\left({r\over r_0}\right),
\end{equation}
and the periodic functions $X_{\pm0}$ and $g_0$ are free
parameters of the general solution. The periodic function $a_0$ is uniquely
determined by $X_{\pm0}$ through the ODE
\begin{equation}
a_0' = {1 \over 2} a_0 \left[(1 - a_0^2) +
a_0^2 (X_{+0}^2 + X_{-0}^2)\right],
\end{equation}
and the the periodic function $\sigma$ and constant $\nu$ are determined
from $a_0$ by integration, as
\begin{equation}
\label{nu}
\nu = \widehat{(1-a_0^2)}_0,
\qquad \sigma = I_0(1-a_0^2),
\end{equation}
where the notation is that of appendix \ref{appC}.

From this asymptotic expansion we see that $a$ and $X_\pm$ are
regular at $t=0$, while $g$ is not and will have to be replaced by another
dependent variable.

We replace $\zeta$ and $\tau$ by $\rho$ as above and $w$ given by
\begin{equation}
\label{w}
w \equiv e^{
(1 + n)(\tau - \rho) + f(\tau) +h(\rho)},
\end{equation}
where $n$ is a constant and $h(\rho)$ and $f(\tau)$ are periodic functions
of one variable, all yet to be determined. Among the dependent variables,
we replace $g$ by
\begin{equation}
\label{defG}
\bar g \equiv \left[1 + n + f'(\tau)\right] \, e^{n (\tau - \rho) + f(\tau)
+h(\rho)} g.
\end{equation}
We now have DSS if the dependent variables $a$, $\bar g$,
$X_+$ and $X_-$ are periodic in $\rho$ at constant $w$, with the same
period $\Delta$ we have determined previously.

The field equations in the new dependent and independent variables are
\begin{eqnarray}
\label{matter2}
X_{\pm,w} && = 
{C_\pm - X_{\pm,\rho} \over - \Gamma w \mp \bar g}, \\
\label{bigG}
\bar g_{,w} && = {\bar g_{,\rho} + (a^2 - 2 + \Gamma) \bar g \over \Gamma w}, \\
\label{a2}
a_{,w} && = \bar g^{-1} C_2, \\
\label{constraint2}
a_{,\rho} && = C_1 + \Gamma w \bar g^{-1} C_2,
\end{eqnarray}
where
\begin{equation}
\label{Gamma}
\Gamma(\rho) \equiv 1 + n - h'(\rho).
\end{equation}
We are dealing again with four evolution equations (now in $w$ instead
of $\zeta$) (\ref{matter2}-\ref{a2}) and one constraint
(\ref{constraint2}). Singular points arise where the denominator of
(\ref{matter2}) or of (\ref{bigG}) vanishes.

We first examine the singular point $w=0$, corresponding to $t=0$.  We
assume that $a$ is at least $C^0$ there, as suggested by the asymptotics,
and takes the value $a_0(\rho)$. If we set $a=a_0(\rho)$ in equation
(\ref{bigG}), this approximate equation admits an exact general
solution, which we can write as
\begin{equation} 
\bar g(w,\rho) = w^{n - \nu \over 1 + n} \
\exp \left\{ {-{n - \nu \over 1 + n} h(\rho)} 
{-I_0(a^2 - 2 + \gamma)(\rho)} 
+{F\left[\rho + {\ln|w| - h(\rho) \over 1 + n}\right]} \right\}
\end{equation}
Here $\nu$ is defined by equation (\ref{nu}), and $F$ is a periodic
function of one variable with period $\Delta$, which serves as the
free parameter of the solution.  We see that $\bar g$ will either blow
up or vanish at $w=0$, unless we impose $n=\nu$. Furthermore $\bar g$
will not be $C^0$ unless $F\equiv 0$. But imposing these two
conditions is equivalent to the one condition
\begin{equation}
\label{b3}
\left[ \bar g_{,\rho} + (a^2 - 2 + \Gamma) \bar g \right]_{w=0} = 0.
\end{equation}
(This is so because the expression $a^2 - 2 + \Gamma$ can only be the
derivative of the periodic function $\ln \bar g$ if its average value
vanishes. This corresponds precisely to $n=\nu$, although $\nu$ does
of course not appear explicitly in the equations.)  This is just the
regularity condition one would have expected from inspecting the
equations, but now we have shown that imposing it actually corresponds
to imposing analyticity.

Before we discuss the other singular points, we use $h(\rho)$ to
identify $\zeta = \zeta_0$ with $w = 1$, thus simplifying the matching
between the two coordinate systems. (We then need numerical
interpolation only in one, not two dimensions). We define the periodic
function $\hat\xi_0$ implicitly in terms of $\xi_0$ and $\zeta_0$ by
the equation
\begin{equation}
\label{hatxi}
\hat\xi_0[\tau + \zeta_0 + \xi_0(\tau)] \equiv \xi_0(\tau). 
\end{equation}
(Numerically, this is solved by iteration.) Next we define
\begin{equation}
\label{hatf}
\hat f(\rho) \equiv f[\rho - \zeta_0 - \hat\xi_0(\rho)].
\end{equation}
This definition implies that $\hat f(\rho)$ and $f(\tau)$ coincide
when restricted to the line $\zeta=\zeta_0$.  Now we fix $h$ as
\begin{equation}
\label{hdef}
h(\rho) \equiv (1 + n) \left[\zeta_0 + \hat\xi_0(\rho)\right] - \hat f(\rho),
\end{equation}
and it can be verified that with this definition $w=1$ for $\zeta=\zeta_0$.
Now let $a_0(\tau) \equiv a(\zeta_0, \tau)$, and define $\hat a_0$ from
$a_0$ in the same way as $\hat f$ from $f$. Proceed similarly for $g$,
$X_+$ and $X_-$. 
$\hat a_0$ and $\hat X_{\pm0}$ now constitute initial data for $a$ and
$X_\pm$ on $w=1$. The initial data for $\bar g$ on $w=1$ can be expressed in
terms of $\hat g_0$ as
\begin{equation}
\label{b4}
\bar g(w=1,\rho) = {\Gamma \over 1 - \hat\xi_0'} \,  e^{\zeta_0 +
\hat\xi_0(\rho) } \, \hat g_0(\rho).
\end{equation}

Now we come back to the other singular points of the equations, namely
where $-\Gamma w \mp \bar g = 0$.  Straightforward calculation shows
that this happens where $dw=0$ is null, just as the denominator of
(\ref{matter}) vanishes where $d\zeta=0$ is null. Either occurrence
indicates a SSH. If we choose $\zeta_0$ large enough, we do not
encounter the past SSH. Why don't we simply choose $\zeta_0=0$,
however?  As one expects, the denominator of the matter equation then
vanishes on the entire line $w=1$, but it vanishes also at two more
lines crossing $w=1$. This is confirmed by the fact that the partial
derivative of the denominator with respect to $w$ vanishes at two
points (per period) on $w=1$. This behavior would give us the
numerical problems we avoided in the old coordinate system by making
the SSH a line of constant $\zeta$. (In contrast, the denominator of
the matter equation in the $(\zeta,\tau)$ coordinate system is
everywhere increasing with $\zeta$, which means that it vanishes only
at $\zeta=0$ and nowhere else.)

We have now formulated the analytic continuation at $t=0$ as a
boundary value problem mathematically quite similar to the one we have
solved above. We consider $a$ as determined by $\bar g$ and $X_\pm$
via the constraint (\ref{constraint2}). The unknowns are the fields
$\bar g$ and $X_\pm$ between $w=1$ and $w=0$, and the function
$\Gamma(\rho)$. The three boundary conditions on the left are the
matching of $\bar g$ and $X_\pm$ to the data at $\zeta = \zeta_0$, and
the one boundary condition on the right is (\ref{b3}). Note that the
unknown functions and constant appear in the field equations only in
the one combination $\Gamma$, and that $\hat g(w=1)$ is determined
from $\Gamma$, $\hat\xi_0$ and $\hat g_0$. This breaks up our
numerical procedure naturally into two steps. In a first step, the
functions $\hat\xi_0$, $\hat a_0$, $\hat g_0$ and $\hat X_{\pm 0}$ are
once and for all determined from the data at $\zeta=\zeta_0$. Then we
vary $\Gamma$ in a relaxation algorithm, until we have solved the
system (\ref{matter2}-\ref{constraint2}) with boundary condition
(\ref{b3}).

We have seen that in the boundary value problem we are only dealing
with one free function $\Gamma(\rho)$ to be determined, while $f$, $h$
and $n$ do not appear explicitly. I have introduced them because they
are useful in deriving an initial guess for $\Gamma$ for use in the
numerical algorithm.  Substituting the asymptotic form (\ref{ginf}) of
$g$ into the definition of $\bar g$ we obtain
\begin{equation}
\bar g \simeq g_0(\tau) \, \left[1 + n + f'(\tau)\right] \, e^{f(\tau) -
\nu \xi_0(\tau) + (n - \nu) (\tau - \rho) + \sigma(\rho) + h(\rho)}
\end{equation}
(Note that, exceptionally, I have mixed coordinate systems, in using $\rho$
and $\tau$ as independent variables.) This is regular at $\tau = \infty$
if $n=\nu$, and if 
\begin{equation}
\label{feqn}
g_0(\tau) \, \left[1 + n + f'(\tau)\right] \, e^{f(\tau) -
n\xi_0(\tau)} = 1.
\end{equation}
This latter condition can be considered as a linear ODE for $\exp
f(\tau)$, given $g_0$ and $n=\nu$. By evolving in coordinates
$(\zeta,\tau)$ to large $\zeta$, and comparing with the asymptotic
form (\ref{ginf}) we can estimate $\nu$ and $g_0$. From there we can
calculate an estimate for $f$ via (\ref{feqn}) and then for $h$ and
hence $\Gamma$ via (\ref{hatxi}, \ref{hatf}, \ref{hdef}).

Now that the problem has been solved in the new variables, we can
calculate $\alpha = a / g$ from (\ref{defG}). It contains a singular
$t$-dependent factor that we can absorb into the (singular)
redefinition $t \to \bar t = t^{1+n} e^{f(\ln t/r_0)}$. The
resulting regular $\bar \alpha$ is
\begin{equation}
\bar \alpha = e^{-n \rho + h(\rho)} \, {a \over \bar g}.
\end{equation}
$n$ and $h$ are given in terms of the now determined function $\Gamma$
by $n=\hat \Gamma_0 - 1$ and $h = - I_0 \Gamma$.  Note that $\bar
\alpha$ is no longer periodic in $\rho$, although the spacetime is
DSS. We have had to use the most general form of the metric
compatible with (\ref{metric}), abandoning the gauge condition
$\alpha=1$ at $r=0$. In fact $\bar \alpha$ is singular at $r=0$, but
then we only use it for a patch around spacelike infinity.

We obtain a clearer picture of the behavior at spacelike infinity by
ignoring the ``wiggles'' in $a$ and $\alpha$. Then they simplify to the
expressions
\begin{equation}
a \sim \sqrt{1 - n}, \qquad \bar\alpha \sim r^{-n},
\end{equation}
which are valid for $r\gg |t|$. The Hawking mass is proportional to
the radius, and the geometry at spacelike infinity is conical.  Our
constant $n$ coincides with ``$1/n$'' in the notation of
\cite{EvCol}. Numerically I find $n \simeq - 0.16$.


\subsection{From $t=0$ to the future self-similarity horizon}


It remains to extend the spacetime all the way to the future SSH, the
future light cone of the singularity at $(t=0,r=0)$, beyond which the
continuation is no longer unique. A priori, the future SSH might itself be
singular, because we have no free parameters left to make it regular. 

At $t=0$, the periodic data $X_\pm(\rho)$ are fixed, and they determine
$a(\rho)$ through the constraint (\ref{constraint2}). $\bar g$ on the other
hand is pure gauge at $t=0$, depending on our choice of $\Gamma$ through
equation (\ref{b3}). We now continue to evolve in $w$ with the same
equations as before, but with a new choice of $\Gamma$. This means that we
introduce a third coordinate system, although one of the same class
(\ref{rho},\ref{w}) as before. This time we choose $\Gamma$ so as to make
the line $w=-1$ the future SSH, which is also the future light cone of the
singularity. Our reason for this is twofold: on one hand we are better
placed to control the vanishing of the denominator in the matter equation
(now for $X_+$ instead of $X_-$) when it takes places on a coordinate line,
on the other hand we want the edge of the domain of dependence of our data
to coincide with a coordinate line, so that we can evolve right up to
it. Our boundary conditions are now the data for $X_\pm$ and $a$ at $w=0$,
the constraint (\ref{b3}) which determines $\bar g$ from $\Gamma$ at $w=0$
(up to a constant factor), and the coordinate condition $\bar g=\Gamma$
(vanishing of the denominator of $X_{+,w}$) at $w=1$. A priori this new
boundary value problem is not well-posed.  We have no freedom left to
impose the vanishing of the numerator of $X_{+,w}$ at $w=1$ as an
additional boundary condition, so that the solution should be genuinely
singular at this point.

To investigate what happens we once more make an analytic
approximation. Let us assume that $X_-$ is at least $C^0$ at $w=-1$, and
takes value $X_{-0}(\rho)$. By definition, as our new coordinate condition,
$\bar g$ takes the value $\Gamma(\rho)$. In consequence $X_+$ drops out of the
constraint (\ref{a2}), and independently of the value of $X_+$, $a$ takes
the value $a_{0}(\rho)$, which is the solution of the ODE
\begin{equation}
\label{a0}
a_{0}' = {1 \over 2} a_{0} (1 - a_{0}^2) + a_{0}^3 X_{-0}^2
\end{equation}
with periodic boundary conditions. The solution exists and is unique.  We
can now calculate $\bar g_{,w}$ at $w=-1$, and hence the linear approximation to
the denominator $-\Gamma w - \bar g$ near $w=-1$. With these expressions in hand
we write out the equation for $X_+$, in the leading terms in both the
numerator and the denominator. We obtain an approximate equation of the
same form as (\ref{approx}), namely
\begin{equation}
X_{+,w} = {A(\rho) X_+ - X_{+,\rho} + C(\rho)
\over (w+1) B(\rho) },
\end{equation}
where the coefficients are
\begin{equation}
A(\rho) = {1\over2}(1-a_0^2) - a_0^2 X_{-0}^2,
\qquad B(\rho) = a_0^2 - 2 + \left(\ln \Gamma\right)',
\qquad C (\rho) = - X_{-0}.
\end{equation}
The exact general solution to the approximate equation is, once more,
\begin{equation}
X_+ = X_+^{{\rm inhom}}(\rho) + X_+^{{\rm hom}}(w,\rho),
\end{equation}
where
\begin{equation}
A X_+^{{\rm inhom}} - X_{+,\rho}^{{\rm inhom}} + C = 0
\end{equation}
and
\begin{equation}
X_+^{{\rm hom}} = (w+1)^{\hat A_0 \over \hat B_0} 
\ e^{I_0 A - {\hat A_0 \over \hat B_0} I_0 B }
\ F\left[\rho + {I_0 B - \ln|w+1| \over B_0}\right]
\end{equation}
As we have no freedom to impose any boundary conditions, $F$ does not
vanish and the solution is not analytic at $w=-1$. But we see that $\hat
B_0 < 0$, while $\hat A_0 < 0$ unless $(a_0-1)$ and $X_{-0}$ vanish
identically. The infinitely oscillating term
therefore vanishes at $w=-1$ as $(w+1)^\epsilon$, where $\epsilon$ is
positive and small.  $X_+$ is therefore $C^0$, although not $C^1$.
This is a remarkable result: The presence of even a small amount of
radiation crossing the future SSH (the component $X_-$) regularizes the
radiation running along the horizon (the component $X_+$), by damping its
oscillations. A similar result was found by Horne in the axion-dilaton
\cite{EHH} and free complex scalar \cite{HE1} (see note added in print) CSS
solutions, using a similar analytic approximation. 

From the field equations it follows that $a$, $\bar g$ and $X_-$ are $C^1$
but not $C^2$ (with respect to $w$, differentiation with respect to $\rho$
is not a problem). As not all second derivatives of the metric exist, one
must ask if the spacetime curvature is finite. In appendix \ref{appG} I
show that all components of the Riemann tensor are in fact $C^0$.

Although the numerical problem is ill-defined because of the presence of an
infinite number of oscillations in $X_+$, I have run a naive relaxation
algorithm on it. The algorithm does in fact converge, and I even see
convergence in the values of $\Gamma$, $a_{0}$ and $X_{-0}$ with decreasing
step size.  I find $\Gamma \simeq 1 + \nu$, and $a_0 \simeq 1$ and $X_{-0}
\simeq 0$. This means that spacetime is approximately flat on the future
SSH, and very little scalar field radiation is crossing it.  $X_+$ however,
oscillates more and more rapidly. These oscillations appear to be at
constant amplitude, and I do not see their eventual decay numerically. This
is consistent with the fact that $\epsilon \ll 1$, so that the decay is
very slow.

In order to make the problem numerically well-defined, in spite of the
solution being singular, one would have to subtract the singular part,
solving for $F$ in another boundary value problem. (This caveat may apply
also to Horne's numerical results \cite{EHH}.) I have not attempted this
nontrivial step, as I do not see an immediate need for quantitative
results. Although the problem is not numerically well-defined as it stands,
I am confident in the qualitative result that the null data at the future
SSH are regular, and nearly, but not quite flat.

Why all this explains in hindsight why the future SSH is, in some sense,
regular, it does not explain why it is nearly flat, that is, why $X_{-0}$
is so small.  I do not see a mechanism in the field equations that would
drive arbitrary data at $w=0$ to very nearly flat space values. And in fact
the SSH tends to be less flat when I slightly perturb $X_+$ or $X_-$ at
$w=0$ (and adjust $a$ accordingly). This indicates that the near-flatness
at the future SSH is a property of the special DSS solution which is
regular at the past SSH. (Incidentally, it is also an argument against
near-flatness being a numerical artifact.)


\section{Linearized perturbations and critical exponent}
\label{sec:pert}


\subsection{The linear eigenvalue problem}

Now we turn to the study of linearized perturbations of the critical
solution, which leave the perturbed solution regular at $r=0$ and
$\zeta=0$. If such a regular perturbation existed while being periodic in
$\tau$, the critical solution would not be isolated. The perturbations must
therefore break the discrete homotheticity, that is, the periodicity in
$\tau$. The coefficients of the linearized equations, however, are
periodic, and therefore the general linear perturbation is a superposition
of terms of the form
\begin{equation}
\delta Z(\zeta,\tau) = \sum_{i=1}^{\infty} C_i e^{\lambda_i\tau}\, \delta_i Z(\zeta,\tau),
\end{equation}
where each $\delta_i Z$ is periodic in $\tau$ with period $\Delta$
(although $\delta Z$ is not!). $Z$ is a shorthand for $(a,g,X_+,X_-)$.  The
exponents $\lambda_i$ and hence the $\delta_i Z$ must be allowed to be
complex even for real $\delta Z$. As the equations are real, they form
complex conjugate pairs, corresponding to sine and cosine oscillations in
$\tau$ with frequency $\Im\lambda_i$. Because the ansatz already contains
all frequencies that are integer multiples of $2\pi/\Delta$, and because
values of $\lambda_i$ come in complex conjugate pairs, we need consider
only $0 \le \Im\lambda_i < \pi/\Delta$.

If the equation for $\delta Z$ is of the form
\begin{equation}
\delta Z_{,\zeta} = A \, \delta Z + B \, \delta Z_{,\tau},
\end{equation}
the equation for $\delta_i Z$ is of the form
\begin{equation}
\delta_i Z_{,\zeta} = (A + \lambda_i B) \, \delta_i Z + B \,
\delta_i Z_{,\tau}.
\end{equation}
This indicates how we obtain the equations for $\delta_i Z$ from the
linearized equations for $\delta Z$.  In the following I denote the
components of $\delta_i Z$ by $\delta_i a$, $\delta_i g$ and $\delta_i
X_\pm$. The equations for the periodic quantities $\delta_i Z$ are then
\begin{eqnarray}
\nonumber
\delta_i  X_{\pm,\zeta} = &&
\Biggl\{
\left[{1\over2} (1-a^2) -a^2 X_\mp^2 \right] \delta_i  X_\pm
-\left(2a^2X_+X_- + 1\right) \delta_i  X_\mp
-aX_\pm\left(1 + 2 X_\mp^2\right) \delta_i  a \\
&& \pm e^{\zeta + \xi_0} \left[ X_{\pm,\tau} \delta_i  g
 + g \left( \delta_i  X_{\pm,\tau} + \lambda_i \delta_i  X_\pm\right)
- (1+\xi_0') X_{\pm,\zeta} \delta_i  g \right] 
\Biggr\} 
\left[1 \pm (1+\xi_0') e^{\zeta + \xi_0} g\right]^{-1} \\
\delta_i  g_{,\zeta} = &&  (1 - a^2) \delta_i  g - 2 a g \delta_i  a \\
\delta_i  a_{,\zeta} = && \left[ {1\over 2}+{3\over
2}a^2\left(X_+^2+X_-^2-1\right) \right] \delta_i  a
+ a^3 \left(X_+ \delta_i  X_+ + X_- \delta_i  X_-\right) \\
\nonumber
\delta_i  a_{,\tau}  = && - \lambda_i \delta_i  a 
 + e^{-(\zeta + \xi_0)} g^{-1} \left[ {3\over2} a^2 (X_+^2 - X_-^2)
\delta_i  a + a^3 (X_+ \delta_i  X_+ - X_- \delta_i  X_-)
-{1\over 2} g^{-1} a^3(X_+^2 - X_-^2) \delta_i  g\right] \\
\label{dela}
&& + (1+\xi_0') \left\{ \left[ {1\over 2} + {3\over 2} a^2 (X_+^2 +
X_-^2 - 1) \right] \delta_i  a + a^3 (X_+ \delta_i  X_+ + X_- \delta_i  X_-)\right\}
\end{eqnarray}

We now consider the boundary conditions for these equations.  At
$\zeta\to-\infty$ we have one free function $\delta_i Y_1(\tau)$ (compare
appendices \ref{appA} and \ref{appB}).  At the boundary $\zeta = 0$, the
denominator of the $\delta_i X_-$ equation vanishes, because it is the same
as that of the $X_-$ equation. We therefore have to impose the vanishing of
the numerator. Imposing the linear constraint as well, we can freely
specify $\delta_i g$ and $\delta_i X_+$ at $\zeta=0$, and calculate from
them $\delta_i a$ and $\delta_i X_-$. Details are given in appendix
\ref{appB}.

We now have three free functions $\delta_i  Y_{1}$, $\delta_i  g_0$ and $\delta_i 
X_{-0}$ at the boundaries. At some intermediate point the three functions
$\delta_i  g$, $\delta_i  X_+$ and $\delta_i  X_-$ will have to match. The fourth,
$\delta_i  a$, will match automatically by virtue of the linearized constraint
(\ref{dela}).  We also have the free constant $\lambda_i$ to solve for. Its
presence is balanced by the fact that because of linearity an overall
factor in the perturbations is arbitrary and has to be fixed in some way.
The $\lambda_i$ are the eigenvalues in a new hyperbolic boundary value
problem, this time a linear one. One would therefore expect them to form a
discrete set.

Details of the numerical method are given in appendix \ref{appE}, and an
error estimate in appendix \ref{appF}. Because of the even-odd symmetry of
the background, perturbations which have the same symmetry decouple from
perturbations with the opposite symmetry, and we can consider them
separately.

In the left half plane of $\lambda_i$, corresponding to modes that grow
towards the singularity $\tau = -\infty$, I have found one perturbation of
the first type. In the following I refer to this mode by $i=1$. It is real,
with $\lambda_1 = -2.674 \pm 0.009$, corresponding to a critical exponent
$\gamma = 0.374 \pm 0.001$. I find no perturbations of the second type in
the left half plane. There is one perturbation, of the first type, at
$\lambda_i = 0$. It is the gauge mode $\delta Z \propto Z_{*,\tau}$,
corresponding to a translation of the background in $\tau$. On the positive
real axis I find perturbations of both types, and presumably there are more
elsewhere in the right half plane.


\subsection{Critical scaling of the black hole mass}
\label{subsec:scaling}


The derivation of the scaling of the black hole mass for near-critical
initial data which I present here was suggested in \cite{EvCol}, and first
made explicit in \cite{Koike} for the CSS case. The DSS case requires a
subtle generalization, and I find a new phenomenon: a ``wiggle'' in the
mass scaling law.

In preparation, let us consider a family of Cauchy data at constant $t$,
namely
\begin{equation}
\label{Ztau}
Z_\tau(r) = Z_*\left(\ln {r\over r_0}, \tau\right) + \epsilon\  \delta_1 Z\left(\ln {r\over r_0}, \tau\right),
\end{equation}
where $\tau$ is the parameter of the family.  Here $Z_*(\zeta,\tau)$ is
the critical solution that we have just constructed, and $e^{\lambda_1
\tau} \delta_1 Z(\zeta,\tau)$ is the one linear perturbation mode that is
growing on small scales, that is, which has negative $\lambda$.  $r_0$ is
the (arbitrary) fixed length scale introduced in (\ref{tauzeta}) and
$\epsilon$ is a fixed small constant, small enough so that the linear
approximation is good initially.  $\ln r/r_0$ is the value the formal
argument $\zeta$ takes. Clearly this family is periodic in $\tau$ with
period $\Delta$.

What happens at late times when these data are evolved in $t$? For one sign
of $\epsilon$, say $\epsilon<0$, we must find dispersion, for the other a
black hole. The key observation \cite{HE2} is that the data depend on $r$
only through the dimensionless combination $r/r_0$, while the evolution
equations themselves have no scale. The entire solution scales as 
\begin{equation}
Z_\tau(r,t)=f\left({r\over r_0},{t\over r_0},\tau\right).
\end{equation}
Therefore we know, without having to rely on the linear approximation, that
the black hole mass, which has dimension length, must be proportional to
$r_0$. More precisely, we have
\begin{equation}
M = r_0 \ e^{\mu(\tau)},
\end{equation}
where $\mu(\tau)$ is an unknown periodic function of period $\Delta$. It
can be calculated numerically by evolving members of the family $Z_\tau$
which span one period in $\tau$.

Now we consider a generic solution. If the initial data are
sufficiently close to criticality, there is a spacetime region in
their evolution where the solution ``echos'', that is, where it is
close to the critical solution.  There the solution is well
approximated by the critical solution plus linear perturbations, as
\begin{equation}
Z(\zeta,\tau) \simeq Z_*(\zeta,\tau) 
+ \sum_{i=1}^\infty C_i(p) \ e^{\lambda_i \tau} 
\delta_i Z(\zeta,\tau).
\end{equation}
The amplitudes $C_i$ of the perturbations depend in a complicated way on
the initial data in general and hence on the parameter $p$ of a given
one-parameter family of initial data.  As $t \to 0$ but while the
perturbations are still small, we can neglect all perturbations but the
growing one, and we have
\begin{equation}
Z(\zeta,\tau) \simeq Z_*(\zeta,\tau) 
+ C_1(p) \ e^{\lambda_1 \tau} 
\delta_1 Z(\zeta,\tau).
\end{equation}
By definition we obtain the precisely critical solution for $p=p_*$, and so
we must have $C_1(p_*) = 0$. Linearizing $C_1(p)$ around $p_*$, we obtain
\begin{equation}
\label{app}
Z(\zeta,\tau) \simeq Z_*(\zeta,\tau) 
+ (p-p_*) \ {\partial C_1(p_*) \over \partial p} \ e^{\lambda_1 \tau} 
\delta_1 Z(\zeta,\tau).
\end{equation}
We define \begin{equation}
\gamma \equiv - {1\over \lambda_1} 
\end{equation}
(as the notation suggests, this
will turn out to be the critical exponent), and
\begin{equation}
\tau_*(p)\equiv \gamma \ln \left({p - p_* \over p_*}\right),
\qquad \tau_1 \equiv \gamma 
\ln \left[{1\over \epsilon} {\partial C_1 \over \partial \ln p}(p_*)
\right].
\end{equation}
Note that $\tau_*(p)$ is only a way of rewriting the ``reduced
parameter'' $(p-p_*)/p_*$, while $\tau_1$ depends on the family of
initial data through the unknown function $C_1(p)$.  $\epsilon$ is
the same as in definition (\ref{Ztau}). If we now fix $\tau =
\tau_*(p) + \tau_1$ in the approximate solution (\ref{app}), we obtain
a $p$-dependent family of Cauchy data, namely
\begin{equation}
Z_p(r) = Z_*\left(\ln {r\over r(p)}, \ \tau_*(p) + \tau_1\right) + \epsilon
\  \delta_1 Z\left (\ln {r\over r(p)}, \ \tau_*(p) + \tau_1\right),
\end{equation}
where 
\begin{equation}
r(p) \equiv r_0 \ e^{\tau_*(p) + \tau_1 + \xi_0[\tau_*(p)+\tau_1]}
\end{equation}
But this is of the form for which we have just calculated the black hole
mass. Therefore we have
\begin{equation}
\label{mass}
M(p) = r(p) \, e^{\mu[\tau_*(p) + \tau_1]} = 
r_0 \left({p-p_*\over p_*}\right)^\gamma 
e^{\tau_1 + (\mu + \xi_0)[\tau_*(p) + \tau_1]}.
\end{equation}
Let us first consider the CSS case. Then $\mu$ and $\xi_0$ degenerate into
constants. We recover the well-known exact scaling of the black hole
mass. The unknown, family-dependent, constant $\tau_1$ corresponds to an
unknown overall factor in the mass.

In the DSS case we find a ``wiggle'' overlaid on the scaling law, unless
the function $(\mu+\xi_0)$ vanishes identically. The period of both $\mu$
and $\xi_0$ is nominally $\Delta$, but $\xi_0$ has only even frequencies,
and so does $\mu$, being based only on the metric coefficients $a$ and
$g$. The real period in $\tau$ of $(\mu+\xi_0)$ is therefore $\Delta/2$,
and hence the period of the wiggle in the directly measured parameter
$\ln(p-p_*)$ is $\Delta/(2\gamma) \simeq 4.61$.

The offset of the wiggle is the same constant $\tau_1$ that already appears
in the overall factor. Given the function $\mu(\tau)$, the black hole mass
is therefore as much determined in the DSS case, namely up to one
family-dependent constant, as in the CSS case, and $\mu(\tau)$ has
the same universal significance as $\gamma$. It would therefore be
interesting to determine $\mu(\tau)$ from evolving (\ref{Ztau}) and to test
the expression (\ref{mass}) against collapse simulations.


\subsection{The free complex scalar field model}


As mentioned already, there is a regular spherically symmetric CSS solution
for the free complex scalar field, that is a complex scalar field with
neither a mass term nor a coupling to an electromagnetic field
\cite{HE1}. Later it was discovered that this solution has not one but
three unstable modes \cite{HE2}, and that the critical solution for the
free complex scalar field is in fact the real DSS solution we have
discussed here (up to a global complex phase)\cite{privcom}. 

The action is, in loose notation, $R+|\partial \Phi|^2$. Writing the
complex scalar $\Phi$ as $\phi+i\psi$, the action becomes $R+(\partial
\phi)^2 + (\partial \psi)^2$, or that of two real scalar fields.  Any
solution of the real scalar field model is therefore also a solution of the
free complex scalar field model.  Furthermore, purely imaginary scalar
field perturbations $\delta \psi$ of any purely real solution $\phi$
decouple from the purely real perturbations $\delta \phi$. This holds
because the stress tensor is the some of a term quadratic in $\phi$ and one
quadratic in $\psi$. The first-order perturbation of this second term
vanishes if the background value of $\psi$ is zero.

In order to obtain all perturbations of the real scalar DSS solution within
the free complex scalar model, I only need to calculate its purely
imaginary perturbations $\delta \psi$ (in addition to the real
perturbations $\delta \phi$ already known). These obey the wave equation on
the fixed background metric with source $\phi_*$, while the corresponding
metric perturbations vanish to linear order. I have checked that all
these modes are damped. With a little extra work I have thus confirmed
perturbatively that the real DSS solution is an attractor of co-dimension
one even in the free complex model. Details will be given elsewhere.


\section{Conclusions}
\label{sec:concl}


\subsection{Critical solutions and matter models}


As I have argued in the introduction, critical solutions of the kind
discussed here play a role in critical collapse similar to the role the
Kerr-Newman solution plays in generic, non-critical, collapse. The crucial
difference is the presence of matter: A critical solution does not have a
horizon, and therefore depends explicitly on the choice of matter. This
reduces the importance of any one such solution. Two other differences are
mainly of technical importance. While the Kerr-Newman solutions are known
in closed form, all critical solutions discovered so far are only known
numerically. Furthermore, CSS solutions, like stationary ones, have
effectively one dimension fewer, but this simplification does not hold for
the DSS ansatz: it only makes all fields periodic in one coordinate. The
labour involved in constructing the real scalar field critical solution,
which is DSS, is therefore much greater than for the other self-similar
solutions found so far, all of which are CSS. There are two reasons for
investing it. On one hand the real scalar field model is the first in which
critical phenomena were observed, and its ``echos'', and the ``wiggle'' in
the mass scaling, are new features that differ from the well-known CSS
case. On the other hand, the real scalar field is a testbed of methods for
dealing with DSS which can now be applied to other DSS critical
solutions. The most interesting of these is probably that of pure gravity
in axisymmetry \cite{AbrEv}.

This raises the fundamental problem of angular momentum and electric
charge in the initial data for gravitational collapse. On one hand,
the resulting black hole can have angular momentum and charge. On the
other, both must be smaller than the mass. So what happens to the
black-hole charge and angular momentum if one tries to fine-tune the
black-hole mass to zero? Clearly the case of rotation is the more
interesting one, but it cannot be treated in spherical symmetry. Only
in one case so far has critical collapse be considered in axisymmetry
\cite{AbrEv}, but unfortunately without angular momentum. 

Another restriction of the matter models which have so far been considered
in the study of critical phenomena is that most of them, in marked
difference from any realistic macroscopic matter, do not introduce a
preferred length scale. (In units where $G=c=1$, this is equivalent to the
absence of dimensionful parameters in the action.) This guarantees the
existence of an exactly self-similar solution. Even in the presence of a
preferred scale, however, a self-similar solution may still be a good
approximation towards the singularity, as $s^2\equiv r^2+t^2 \to 0$. This
is the case for one known exception, namely when a mass-term $m^2 \phi^2$
is added to the scalar field action. Then the DSS solution we have
constructed here is a good approximation for $s \ll m^{-1}$, simply because
$\phi$ is bounded while $\partial \phi$ is not. In collapse simulations, it
is found that the Choptuik solution is in fact the attractor for the
massive scalar field \cite{Chop2}. It remains to be investigated how other
matter models introducing a scale react to the attempt to make small black
holes.  A very recent example is the Einstein-Yang-Mills system, which has
one intrinsic scale, and which shows both a mass gap, and critical mass
scaling, for different classes of initial data \cite{EYM}. A rather
different, and interesting, one is the attempt to consider semi-classical
Einstein equations, with the new scale being the Planck length
\cite{ChibaSiino}.


\subsection{Other self-similar scalar field solutions}


In this paper, I have considered a scalar field $\phi$ which is (a) bounded
on the entire spacetime ($\kappa=0$), (b) real, and (c) discretely
self-similar (DSS). These conditions are suggested by the universal
attractor that Choptuik \cite{Chop} found in critical collapse of a real
scalar field, and they are justified by the fact that the unique solution
of the resulting eigenvalue problem coincides with Choptuik's attractor to
numerical precision \cite{PRL}. I now review some related work on
self-similar scalar field solutions departing from one or the other of
these assumptions.

CSS real scalar field solutions have been studied in \cite{Brady}. The
solution for $\kappa=0$ can be derived in closed form and was apparently
first published in \cite{Roberts}, then rediscovered in the context of
critical phenomena in \cite{BradyCQG,Oshiro}. CSS arises as a degenerate
case of our formalism when $a$, $g$ and $X_\pm$ are not only periodic in
$\tau$, but do not depend on $\tau$ at all. In our notation, continuous
self-similarity implies that $Y_1(\tau)$ in equation (\ref{Y_1}) is a
constant, namely $Y_1 = \kappa$. (This is incompatible with the even-odd
symmetry I have assumed.)

Brady \cite{Brady} has shown that for all $\kappa$ in the CSS case,
boundedness at the past CSS is automatic, in marked contrast to the
DSS case. As in the DSS case, continuation past the SSH is not
unique. Brady has considered the one-parameter family of possible
continuations (for each value of $\kappa$). Unfortunately, it is not
clear which of his continuations is the analytic one. In the DSS case,
I have investigated only the analytic continuation at the past SSH,
because the past SSH is in the Cauchy development of regular data when
we think of it as arising as an attractor in collapse simulations, and
therefore I believe it should be analytic. Of Brady's results I
review here only the special case $\kappa=0$, because it appears to
be most closely related to our case. It is qualitatively different
from $\kappa\ne 0$, in that no CSS solution with a regular center
exists, except for flat empty space. There is a one-parameter family
of solutions with a singular center $r=0$. One branch of $r=0$ is
always timelike and has negative mass, the other is either timelike
with negative mass (called subcritical solutions in \cite{BradyCQG}),
or spacelike with positive mass, and in the latter case its is
preceded by an apparent horizon (supercritical solutions). In either
case both the past and the future SSH carry flat-space null data, and
one can therefore replace the negative mass part of the solution with
flat space in both the past and future light cone of the
singularity. Subcritical solutions thus pieced together are qualitatively
like the maximally extended Choptuik solution: $r=0$ is a regular
center, except at the single point $(r=0,t=0)$, which is a naked
singularity. Both the past and the future SSH are regular and
flat. (In the Choptuik solution the past SSH is not flat, and the future
SSH is only nearly flat.)

What would happen if we allowed $\kappa\ne 0$ in DSS, while still imposing
regularity at the center and on the past SSH? Because of mode coupling,
this would mean giving up the even-odd symmetry, and doubling the
(infinite) number of degrees of freedom in the boundary value
problem. Therefore, one might obtain an infinity of new solutions, or no
new solutions at all. If any new solutions exist, one would wonder next why
it is only the one with $\kappa=0$ that serves as a universal attractor. If
there is a family of such solutions continuous with the Choptuik solution,
that question would be even more acute. I leave these questions to future
work.

A CSS solution (with $\kappa=0$ assumed implicitly) has also been
constructed for the free complex scalar field \cite{HE1}. In this solution,
only the metric and the scalar field modulus are CSS, but the complex phase
of the scalar field is $i\omega\tau$ for some constant $\omega$, so that
the complex scalar field might be considered DSS with (in our notation)
$\Delta = 2\pi/\omega$. Not surprisingly, our solution shares more features
with this one than with the CSS real scalar field models: There is a unique
solution with a regular center and regular past SSH, and the null data on
the future SSH are for nearly flat space. The same qualitative picture was
found for the CSS solution with axion-dilaton matter \cite{EHH}. It is
remarkable that now three different self-similar solutions are known which
are nearly, but not quite, flat at the future SSH. The Roberts solution is
clearly the limiting case, with a flat SSH (but consequently a scalar field
that is not $C^0$!), and may be of help in finding an explanation.


\subsection{Perturbations, universality, and ``renormalisation''}


I have found exactly one unstable mode of the critical solution. The
picture of an attractor of co-dimension one is thus confirmed
perturbatively. The Lyapunov exponent of $\simeq -2.674\pm 0.009$ gives
rise to the value $\gamma \simeq 0.374 \pm 0.001$ of the critical
exponent. This value agrees with the most precise value obtained from
collapse calculations \cite{HamStew}, which is given as $0.374$.
Allowing for complex scalar field perturbations around the real Choptuik
solution does not add any unstable modes. The Choptuik solution is therefore
an attractor of co-dimension one even for the complex scalar field.

It would be interesting to run collapse calculations for a one-parameter
family of matter models, such as the $p=k\rho$ family \cite{Maison} or the
non-linear sigma model \cite{HE3}, where the Lyapunov exponents of the
perturbations change continuously with the parameter. One might thus be
able to find parameter regions where two equally strong attractors coexist,
and new interesting phenomena would arise. It would also be extremely
interesting if one could find a way of calculating or estimating the number
of unstable modes of a given self-similar solution other than by
constructing the solution and all its perturbations explicitly.

Another point should be addressed in future theoretical work. As I
mentioned in the introduction, the phase space picture of universality,
although apparently correct in some aspects, is strictly speaking wrong
because the same spacetime corresponds to many different trajectories in
superspace, according to the way it is sliced. At the very least one should
be able to derive a universal geometrical prescription fixing the lapse and
shift from Cauchy data in superspace, such that the intuitive phase space
flow is a realized as a Hamiltonian flow. Furthermore, this Hamiltonian
flow should admit a geometric interpretation as a scaling, or
``renormalisation group'', flow.  The evolution in the time variable $\tau$
(at constant $\zeta$) does in fact go from one set of Cauchy data to the
same, only changed in overall scale.

Clearly the idea of approximate scale invariance is at the core of
renormalisation group ideas and methods. Furthermore, the calculation of
$\gamma$ as given in section \ref{subsec:scaling} is identical with the
calculation of the critical exponent governing the divergence of the
correlation length given in any textbook on critical phenomena in
statistical mechanics, for example \cite{Yeomans}. As far as I can see,
however, the ``critical phenomena'' analogy is not with critical phenomena
in a system which is described by a partition function (such as a system in
thermal equilibrium or a quantum field). There is neither a nonvanishing
Hamiltonian, nor an inverse temperature $\beta$, or quantum of action
$\hbar$, such that one could construct a weight on phase space. 

The analogy seems to be rather with the application of renormalisation
group methods to deterministic PDEs \cite{Goldenfeld}. This takes up ideas
of Barenblatt's \cite{Barenblatt} of generalizing a self-similar ansatz
such as $f(r^2/t)$ for a PDE such as a generalized diffusion equation to
$t^\alpha f(r^2/t)$. Here, $\alpha$ is a non-integer ``anomalous
dimension'' that cannot be determined by dimensional analysis. But the
factor $t^\alpha$ gives rise to terms proportional to $\alpha$ in the
equation for $f$, and this equation will admit regular solutions only for
certain values of $\alpha$: one has a nonlinear eigenvalue problem. 

In the critical collapse case we see this twice. In our ansatz for the
perturbations we clearly have an explicit factor $t^\lambda$. But $\Delta$
has the same function in the background solution! There we expand all
fields, which are periodic in $\tau$, in terms of
$e^{in\omega\tau}f_n(\zeta)$, where $\omega\equiv 2 \pi /\Delta$. In
consequence we obtain terms proportional to $\omega$ in the equations for
the expansion coefficients $f_n(\zeta)$. The CSS case corresponds to these
terms being absent. We may therefore consider the DSS ansatz as a
generalization of the CSS ansatz parallel to the anomalous dimension in
Barenblatt's examples, and consider $i\omega$ a complex anomalous
dimension. These parallels certainly put the investigation of critical
gravitational collapse into a wider context, and may yet give rise to new
predictions or a simplified theory.


\acknowledgments

It is a pleasure to thank Pat Brady, Matt Choptuik, Gary Gibbons, Jim
Horne, Jos\'e Mar\'\i a Mart\'\i n Garc\'\i a, Alan Rendall and John
Stewart for helpful discussions or suggestions. This work was supported by
the Ministry of Education and Science (Spain).


\appendix


\section{Background boundary conditions}
\label{appA}


$\zeta=-\infty$ and $\zeta=0$ are singular
points of the equations. The boundary conditions are therefore
implemented by expanding the field equations in powers of $e^\zeta$
around $\zeta=-\infty$ and of $\zeta$ around $\zeta=0$. The resulting
equations are given below. 

The regularity conditions at $\zeta\to -\infty$ can be solved in closed
form. We expand in powers of $e^{\zeta}$, as
\begin{equation}
a(\zeta,\tau) = a_0(\tau) + a_1(\tau) e^\zeta + a_2(\tau) e^{2\zeta} +
...
\end{equation}
It turns out to be more natural to expand $X$ and $Y$ instead of $X_+$ and
$X_-$.  As discussed in section \ref{sec:critsol}, we impose $a_0=1$ and
$g_0=1$. Expanding the field equations, we find that the two (periodic)
functions $Y_1(\eta)$ and $\xi_0(\eta)$ can be chosen freely.  Their
significance here is the following. $\xi_0$ parameterizes the class of
spacetime coordinates we use, while a combination of the two contains free
boundary data for the scalar field $\phi$ \footnote{This quantity was
called $Y_0(\tau)$ in \cite{PRL}.}:
\begin{equation}
\label{Y_1}
e^{-\xi_0(\tau)} Y_1(\tau) = \sqrt{2\pi G}\left.{\partial \phi \over \partial (\ln t)}
\right|_{r=0}.
\end{equation}
The other nonvanishing
expansion coefficients up to order $e^{3\zeta}$ in $X$ and $Y$ and order
$e^{2\zeta}$ in $a$ and $g$ are
\begin{eqnarray}
a_2 = && - g_2 = {1 \over 3} Y_1^2, \\
X_2 = && {1 \over 3} e^{\xi_0} \left(Y_1' - (1 + \xi_0') Y_1\right), \\
Y_3 = && - {2 \over 3} Y_1^3 +  e^{\xi_0} \left[ 
{1 \over 2} X_2'- (1 + \xi_0') X_2\right], 
\end{eqnarray}
while the other coefficients vanish. These coefficients were
calculated from $a_0 = g_0 = 1$ and the evolution equations alone, but
they also obey the constraint order by order.

Regularity conditions at the boundary $\zeta=0$ give rise to an ODE system. 
We again make a power-law ansatz of the form
\begin{equation}
a(\zeta,\tau) = a_0(\tau) + a_1(\tau) \zeta + a_2(\tau) \zeta^2 + ...
\end{equation}
Here we find that suitable fields to expand in are $a$, $X_+$, $X_-$ and,
instead of $g$, the quantity
\begin{equation} D = (1 + \xi_0') e^{\zeta + \xi_0} g.
\end{equation}
Its evolution equation is easily seen to be
\begin{equation}
\label{ddot}
D_{,\zeta}=D(2-a^2).
\end{equation}
The expansion coefficients can be calculated recursively, by the
solution of first-order ODEs, if we use $X_{+0}(\tau)$ and
$\xi_0(\tau)$ as the free parameters. Our coordinate condition
(\ref{coordcond}) is simply
\begin{equation}
D_0=1.
\end{equation}
We note that $X_-$ drops out of the constraint (\ref{constraint}) to
leading order, which is
\begin{equation}
\label{ODE0}
a_0'=(1+\xi_0') a_0 \left[{1\over
2}+a_0^2\left(X_{+0}^2-{1\over2}\right)\right].
\end{equation}
Given $X_{+0}$, this is a nonlinear, inhomogeneous, first-order ODE for
$a_0$. The one integration constant is fixed uniquely by the requirement
that the solution be periodic.  The regularity condition (\ref{regularity})
gives us another ODE, this time linear, for $X_{-0}$:
\begin{equation}
\label{ODE1}
X_{-0}'+{(1+\xi_0')}\left[
a_0^2\left({1\over2}+X_{+0}^2\right)-{1\over2}\right]X_{-0}+(1+\xi_0')X_{+0}=0.
\end{equation}
We can now calculate algebraically, 
\begin{eqnarray}
D_1=&&2-a_0^2,\\
a_1=&&{1\over 2}a_0\left[
(1-a_0^2)+a_0^2\left(X_{+0}^2+X_{-0}^2\right)\right],\\
X_{+1}=&&{1\over2}\left[-a_0^2X_{+0}\left({1\over2}+X_{-0}^2\right)
+{1\over2}X_{+0}-X_{-0}+ (1+\xi_0')^{-1}X_{+0}'\right].
\end{eqnarray}
For $X_{-1}$ we obtain again a linear ODE,
\begin{eqnarray}
\nonumber 
&&X_{-1}'+ (1+\xi_0')
\left[-{5\over2}+a_0^2\left({3\over2}+X_{+0}^2\right)\right] X_{-1}\\
&&+{(1+\xi_0')}\left[2a_0a_1X_{-0}\left({1\over2}+X_{+0}^2\right)
+2a_0^2X_{-0}X_{+0}X_{+1}+X_{+1}\right]+(2-a_0^2)X_{-0}'=0.
\label{ODE2}
\end{eqnarray}

To quadratic order, we have three algebraic expressions and one more
linear ODE, which we do not give here. (We need the previous equations
to determine $X_{-1}$, which will be needed in equation (\ref{delXm1}) below.)
We have used explicitly only the zeroth order of the constraint
(\ref{constraint}), but the first and second orders are
satisfied identically as expected. 

Using these expansions we can calculate Cauchy data for a very large and a
very small negative value of $\zeta$, thus avoiding the vanishing of
$X_\pm$ at $\zeta=-\infty$ and the breakdown of the Cauchy scheme at
$\zeta=0$.


\section{Boundary conditions for the perturbations}
\label{appB}


The perturbed boundary conditions at $\zeta \to -\infty$ in terms of
the free perturbation $\delta_i  Y_1(\tau)$ are
\begin{eqnarray}
\delta_i  a_2 = && - \delta_i  g_2 = {2 \over 3} Y_1\, \delta_i  Y_1, \\
\delta_i  X_2 = && {1 \over 3} e^{\xi_0} \left(\delta_i  Y_1' + (\lambda_i - 1
- \xi_0') \delta_i  Y_1\right), \\
\delta_i  Y_3 = && - 2 Y_1^2 \delta_i  Y_1 + e^{\xi_0} \left[ 
{1 \over 2} \delta_i  X_2'+ ({1 \over 2} \lambda_i -1 - \xi_0') \delta_i  X_2\right], 
\end{eqnarray}

At $\zeta=0$ the perturbed constraint (\ref{dela}) simplifies
and no longer contains $\delta_i  X_-$:
\begin{equation}
\label{da0}
\delta_i  a_{0,\tau} + \lambda_i \delta_i  a_0 = (1+\xi_0') \left\{
\left[{1\over2} + {3\over2} a_0^2 (2X_{+0}^2 - 1) \right] \delta_i  a_0 +
2a_0^3 X_{+0} \delta_i  X_{+0} - {1\over2} g_0^{-1} a_0^3 (X_{+0}^2 -
X_{-0}^2) \delta_i  g_0\ \right\}.
\end{equation}
We can specify $\delta_i X_{+0}$ and $\delta_i g_0$ freely and solve this
equation for $\delta_i a_{+0}$. The average value of the coefficient of
$\delta_i a_0$ in equation (\ref{da0}) is $\lambda_i + 1$. When this
vanishes, the equation has no solution (with periodic boundary conditions),
as discussed in the appendix on Fourier methods. This means that for
$\lambda_i=-1$, there are no perturbations that are regular on
$\zeta=0$. In the $\lambda$ plane this gives rise to a pole at
$\lambda=-1$, see appendix \ref{appE}.  (To calculate the average value of
the coefficient, we note that the background constraint (\ref{constraint}),
evaluated at $\zeta=0$ with the boundary condition (\ref{coordcond}),
reduces to
\begin{equation}
\left(\ln a_0\right)_{,\tau} = (1+\xi_0')\left[{1\over2}(1-a_0^2) +
a_0^2 X_{+0}^2\right].
\end{equation}
As the left-hand side is the derivative of a periodic function, the
right-hand side has vanishing average.)

The vanishing of the numerator of the $\delta_i X_-$ equation is an ODE
that can be solved for $\delta_i X_{-0}$, namely
\begin{eqnarray}
\label{delXm1}
\nonumber
&& \delta_i  X_{-0}' 
+ \left[\lambda_i - (1+\xi_0') \left({1 \over 2} (1 - a_0^2)
- a_0^2 X_{+0}^2 \right) \right] \delta_i  X_{-0} 
+ (1+\xi_0') \Bigl[a_0 X_{-0} \left(1 + 2 X_{+0}^2
\right) \delta_i  a_0 \\
&& + \left((1+\xi_0')^{-1} X_{-0}' - X_{-1}\right) g_0^{-1} \delta_i  g_0
+\left(2 a_0^2 X_{+0} X_{-0} + 1 \right) \delta_i  X_{+0} \Bigr] = 0.
\end{eqnarray}
This equation in turn has no solution if the coefficient of $\delta_i
X_{-0}$ has vanishing average. This is the case if $\lambda_i$ equals
the average of $(1+\xi_0')(1-a_0^2)$, which numerically has value
$\simeq -0.385$. This gives rise to another pole.

We do not need to expand the perturbations away from $\zeta = 0$, because
in the discretisation (\ref{discretized}) of the linearized equations we do
not need to evaluate the $\zeta$-derivatives of the perturbations at
$\zeta=0$.


\section{Pseudo-Fourier method}
\label{appC}


The discrete Fourier transform of the $N$ complex numbers $f_n$ is
defined by
\begin{equation}
\hat f_k = {1\over N} \sum_{n=0}^{N-1} f_n e^{{2 \pi i k n \over N}},
\end{equation}
and its inverse is
\begin{equation}
f_n =  \sum_{k=0}^{N-1} \hat f_k e^{-{2 \pi i k n \over N}}.
\end{equation}
The motivation of this definition is of course that the $f_n$ are to
represent the values of a smooth periodic (complex) function $f(x)$ at
$N$ equidistant points over one period.  The essential idea of
pseudo-spectral (here: pseudo-Fourier) methods is to carry out
algebraic operations pointwise on the $f_n$, and integration and
differentiation on the $\hat f_k$, switching from one to the other
with a Fast Fourier Transform algorithm.  A detailed description of
pseudo-spectral methods can be found in \cite{Canuto}. I give here
only the technical information necessary to specify my numerical method.

We shall need to define various operations on the Fourier components
$\hat f_k$ which are to represent operations on $f(x)$. To do this in
a consistent way, we have to start from a {\it definition} of $f(x)$
in terms of the $\hat f_k$. We choose 
\begin{equation}
\label{f(x)}
f(x) = \hat f_0 + \sum_{k=1}^{{N\over 2}-1} \left(
\hat f_k e^{-{2 \pi i k \over \Delta} x}
+ \hat f_{N-k} e^{{2 \pi i k \over \Delta} x} \right)
+ f_{N \over 2} \cos\left({N\pi\over\Delta} x\right),
\end{equation}
where $\Delta$ is the period of $f(x)$.  This expression obeys $f(x_n)=f_n$
for $x_n=n\Delta/N$, but this requirement alone does not uniquely define
it. We can now {\it derive} how any operation on the fictitious function
$f(x)$ is represented as an operation on the $\hat f_k$, for example
interpolation of $f(x)$ to arbitrary values of $x$, differentiation or
integration. Let us begin with differentiation. The corresponding operator
$D$ is given by
\begin{eqnarray}
\nonumber
\left(D \hat f\right)_k && = - {2 \pi k \over \Delta} \hat f_k, \quad 0 \le k \le
N/2 - 1, \\ 
\nonumber
\left(D \hat f\right)_{N/2} && = 0 \\ 
\nonumber
\left(D \hat f\right)_k && = - {2 \pi (k - N)
\over \Delta} \hat f_k, \quad N/2 + 1 \le k \le N - 1. \\
\end{eqnarray}
We see that on differentiation the high-frequency cosine, which takes
values of $\pm 1$ on the $x_n$, is annihilated, because the
high-frequency sine takes the value $0$ on the $x_n$.
For the purpose of integration we define a more general operator,
namely $I_\lambda$ given by
\begin{eqnarray}
\nonumber
\left( I_\lambda \hat f\right)_0 && = {\hat f_0 \over \lambda}, \quad\lambda \ne 0,
\\
\nonumber
\left( I_\lambda \hat f\right)_0 && = 0, \quad \lambda = 0, \quad \\
\nonumber
\left( I_\lambda \hat f\right)_k && = {1 \over \lambda - 2 \pi k /\Delta} \hat f_k, \quad 1 \le k \le
N/2 - 1, \\ 
\nonumber
\left( I_\lambda \hat f\right)_{N/2} && = {\lambda \over \lambda^2 - (\pi N /
\Delta)^2} \hat f_{N/2} , \\ 
\nonumber
\left( I_\lambda \hat f\right)_k && = {1 \over \lambda - 2 \pi (k - N) /\Delta} 
\hat f_k, \quad N/2 + 1 \le k \le N - 1. \\
\end{eqnarray}
Clearly, for $\lambda = 0$ and $\hat f_0 = 0$, this is simply integration,
with the integration constant fixed so that the integral has vanishing
zero-frequency component. If $f(x)$ has a nonvanishing average
(zero-frequency part) $\hat f_0$, then its principal function has a part
$\hat f_0 x$ and is no longer periodic. In this case we enforce periodicity
by setting $\hat f_0$ equal to zero. When $\lambda \ne 0$ however, even for
$\hat f_0 \ne 0$, $I_\lambda f$ solves the ODE
\begin{equation}
\left(I_\lambda f\right)' + \lambda\, I_\lambda f = f,
\end{equation}
with the integration constant chosen such that $I_\lambda f$ is itself
periodic. 
With the help of this definition we can write the
unique periodic solution $f$ of
\begin{equation}
\label{inhom}
f' + g f + h = 0
\end{equation}
in closed form, namely as
\begin{equation}
f = - e^{-I_0 g} I_{\hat g_0} \left( e^{I_0 g} h \right).
\end{equation}
This is of course only the standard use of an integration factor,
written in Fourier components, and so that it obeys periodic
boundary conditions. The expression diverges as $\hat g_0 \to 0$, and
indeed the equation (\ref{inhom}) has no solution with {\it periodic}
boundary conditions for $\hat g_0 = 0$. All the ODEs we have to solve
are of the form (\ref{inhom}) except the constraint
(\ref{constraint}), which is nonlinear, of the form
\begin{equation}
2 a' = g a + h a^3.
\end{equation}
Fortunately, it can be reduced to the linear form (\ref{inhom}) by the
substitution $f=a^{-2}$.

We also need to define an operation that doubles $N$ while representing the
``same'' function $f(x)$, and its inverse. These operations will be
required for ``dealiasing'' -- doubling the $f_k$ before going to the
$f_n$, carrying out the necessary algebraic operations on the doubled $f_n$
and going back to the $f_k$, then halving the $f_k$ and thus throwing away
high frequency noise. With our definition of $f(x)$, doubling means
splitting the high frequency cosine. The doubling operation $B$ from $N$ to
$2N$ components is
\begin{eqnarray}
\nonumber
\left(B \hat f\right)_k && = \hat f_k, \quad 0 \le k \le N/2-1, \\
\nonumber
\left(B \hat f\right)_{N/2} && = (1/2) \hat f_{N/2}, \\
\nonumber
\left(B \hat f\right)_k && = 0, \quad N/2 + 1 \le k \le 3N/2-1, \\
\nonumber
\left(B \hat f\right)_{3N/2} && = (1/2) \hat f_{N/2}, \\
\nonumber
\left(B \hat f\right)_k && = \hat f_{k-N}, \quad 3N/2 + 1 \le k \le 2N-1. \\
\end{eqnarray}
The halving operation $S$ from $N$ to $N/2$ components is
\begin{eqnarray}
\nonumber
\left(B \hat f\right)_k && = \hat f_k, \quad 0 \le k \le N/4-1, \\
\nonumber
\left(B \hat f\right)_{N/4} && = \hat f_{N/4} + \hat f_{3N/4}, \\
\nonumber
\left(B \hat f\right)_k && = \hat f_{k+N/2}, \quad N/4+1 \le k \le N/2-1. \\
\end{eqnarray}

I stress once more that the definitions of $D$, $I_\lambda$, $B$ and
$S$ follow uniquely from the definition (\ref{f(x)}). With our choice,
$D$ and $I_0$ are not the inverse of one another because of the
treatment of the high-frequency cosine. $SB$ is the identity operator,
while $BS$ is a smoothing operator. Our choice of definition is
motivated by the requirement that none of the operations mix the real
and imaginary part, that is $Df$, $I_\lambda f$, $Bf$ and $Sf$ are
all real if $f$ is real. I require this because I typically encode two
independent real functions as the real and imaginary parts of one complex
function.

In a previous code \cite{PRL}, I did not use one consistent definition of
$f(x)$ in the Fourier discretisation. More seriously, the discretisation
mixed real and imaginary parts in some places even where the real and
imaginary part represent two unrelated real functions. I have therefore
replaced it by the one described above. This results in a small change in
the numerical results for the same equations discretized on the same grid,
in particular a slightly changed value of $\Delta$.


\section{Numerical methods for the critical solution}
\label{appD}


I use Fourier transformation to transform $1+1$ hyperbolic PDEs into an ODE
system, and ODE boundary conditions into algebraic boundary
conditions. I use the assumption that $X_\pm$ are odd while $a$, $g$ and
$\xi_0$ are even in $\tau$ to encode the variables $X_\pm$, $g$ and $\xi_0$
as a single complex one, as $\xi_0+ig+X_++iX_-$. The discrete Fourier
transform of this function constitutes my independent variables. Let us
assume that the information contained in them is to correspond to a
sampling of $X_\pm$, $g$ and $\xi_0$ at $N$ points each. Then the total
number of complex variables is also $N$, or $2N$ real variables. Typically
I use $2N=128$. Out of these $2N$ variables, I set the one combination
corresponding to the fundamental sine mode of $\xi_0$ equal to zero,
storing $\Delta$ in its place. As mentioned in the main text, this fixes
the translation invariance in $\tau$ of the problem and balances the
degrees of freedom against the constraints. In the analytic continuation
problem the complex variable function is $\Gamma+iG+X_++iX_-$. For the
linearized perturbations it is $i \delta_i g + \delta_i X_+ + i \delta_i
X_-$, plus the constant $\lambda_i$, corresponding to $3N/2+1$ real
variables. The linear perturbations are normalized by an additional
condition at the right boundary, namely that the root-mean-square of
$X_{+0}$ be $1$.

$\xi_0$ and $\Delta$ are formally part of the dependent variables of the
background problem, although with vanishing $\zeta$-derivative. In the
perturbation problem they are supplied as external fixed parameters,
together with the background solution.  In each case, $a$ or $\delta_i a$
is reconstructed at each step from the other three fields by the solution
of the constraint. This solution can be given in closed form in Fourier
components, as explained in appendix \ref{appC}. Previously, we have
doubled the number of Fourier components, transformed back to real space,
and separated the real and imaginary and the even and odd parts to obtain
our four fields sampled at $2N$ points each. At the same time we have
calculated the necessary $\tau$-derivatives. Now we calculate the necessary
algebraic expressions pointwise in the $2N$ sampling points, put the result
back together in complex form, transform back, and halve the number of
Fourier components.

The ODE system $df/d\zeta = F(f,\zeta)$ is discretized in a standard
way as
\begin{equation}
\label{discretized}
f_{n+1} - f_n = (\zeta_{n+1}-\zeta_n) \, F\left[
{f_{n+1}+f_n \over 2}, \, {\zeta_{n+1}+\zeta_n \over 2} \right].
\end{equation} 
To solve this system of algebraic equations to machine precision together
with algebraic boundary conditions on two sides, I use a standard
relaxation algorithm \cite{Press}.  If I have to integrate these equations
forward (that is, with boundary conditions (``initial data'') only on one
side), I use a second-order Runge-Kutta step as a first guess for
$f_{n+1}$, and then solve the discretized equations to machine precision
using Newton's method. On one hand this is necessary because an explicit
algorithm is unstable for these equations even for very small time
steps. On the other I want the equations to be discretized in the same way
as in the relaxation algorithm when I continue the solution from $\zeta=0$
to large $\zeta$.

Schematically, $F(f,\zeta) = N(f,\zeta) / D(f,\zeta)$. As discussed in the
main text, the SSH announces itself by $D=0$. As a boundary condition we
impose that $D=0$ at $\zeta=0$ for all $\tau$. We then impose $N=0$ as a
second boundary condition at $\zeta=0$. $\zeta=0$ is itself the last grid
point. We see from (\ref{discretized}) that $F$ is never evaluated at
$D=0$. There is therefore no need to treat the singular point $\zeta=0$ in
a special way. If we want to shoot away from $\zeta=0$, however, for
example in order to prime the relaxation algorithm, we need to expand the
field equations in powers of $\zeta$. This is done in appendix \ref{appA}.

The other set of boundary conditions are imposed at
$\zeta=-\infty$. We do this by expanding in powers of $\exp\zeta$ (see
appendix \ref{appA}).


\section{Numerical treatment of the perturbations}
\label{appE}


I have studied the perturbations with two largely independent codes. One
code assumes that $\lambda_i$ is real and that the perturbations have the
same symmetry ($X_\pm$ odd and $a$ and $g$ even) as the background. A
second code allows for complex $\lambda_i$ and perturbations, and the
perturbations either having the same symmetry as the background, or the
opposite symmetry. The two kinds of perturbations decouple because of the
background symmetry, and can be considered separately.
 
I used the following discretisation of the linearized version $d\delta
f/d\zeta = \partial F(f,\zeta) / \partial f \cdot \delta f$ of the ODE
system $df/d\zeta = F(f,\zeta)$:
\begin{equation}
\delta f_{n+1} = \left(1 - {h\over 2} {\partial F \over \partial f}\right)^{-1}
\left(1 + {h\over 2} {\partial F \over \partial f}\right)\cdot
\delta f_n
\end{equation}
The matrix $\partial F(f,\zeta) / \partial f$ is extracted in Fourier
components, by varying one Fourier component of $\delta f$ at a time in the
linearized derivatives $\partial F(f,\zeta) / \partial f \cdot \delta f$.
Its coefficients are evaluated on the background taken at the midpoint
between $f_n$ and $f_{n+1}$, and $h\equiv\zeta_{n+1}-\zeta_n$. This scheme
is explicitly linear, implicit (and stable), and second-order accurate.

For finding the eigenvalues $\lambda_i$, both codes generalize ideas
already used in similar calculations \cite{HE2}, using the linearity of the
equations.  All possible linear perturbations compatible with the boundary
conditions at either $\zeta=0$ or $\zeta=-\infty$ are evolved together to a
midpoint, say $\zeta = -1$. One obtains a complex square matrix d(mismatch
at the midpoint) / d(free parameters at the endpoints). As a test, I have
varied the real and imaginary part of the free parameters
independently. The Cauchy-Riemann equations (with respect to the free
parameters) are obeyed without any numerical error.

This matrix, say $A$, depends on $\lambda_i$. If for a given $\lambda_i$ a
linear perturbation exists which obeys all boundary conditions at both
endpoints, $\det A=0$, and this perturbation is the eigenvector with
eigenvalue zero. We therefore have to search for zeros of $\det
A(\lambda_i)$ in the complex plane. Because $\lambda_i^*$ does not appear
in the equations, $\det A(\lambda_i)$ is holomorphic. Complex conjugation
of $A$ corresponds to a certain interchange of both rows and columns (the
interchange of positive and negative frequencies in $\tau$), and therefore
$\det A(\lambda_i^*)=\det A(\lambda_i)^*$. I have checked both symmetries
numerically. I find a relative numerical error in the Schwarz reflection
principle of $10^{-10}$ and in the Cauchy-Riemann equations (with respect
to $\lambda_i$) of $10^{-9}$.

It is sufficient to consider the strip $0 \le \Im \lambda_i <
\Delta/\pi$. Because $\det A(\lambda_i)$ is holomorphic, we can use the
well-known contour integral $\int A'/A\,d\lambda_i = 2\pi i (N_z - N_p)$ to
count the number of zeroes minus the number of poles within the
contour. Apart from the perturbations discussed in section \ref{sec:pert},
I find a simple pole at $\lambda_i = - 1$, for modes with the same symmetry
as the background, and a simple pole at $\lambda \simeq -0.385$ for modes
with the opposite symmetry. Their origin is discussed in appendix
\ref{appB}.

Once I have found a zero of $\det A(\lambda_i)$, I determine the zero
eigenvector, and hence the corresponding perturbation
field. (Numerically I find the eigenvector by singular value
decomposition.) To refine this result, and to provide an independent
check, I use it as the starting point of a relaxation algorithm. In
this algorithm, $\lambda_i$ is one of the dependent variables, and its
presence is balanced by a boundary condition fixing the norm of the
linear perturbation. I have implemented this algorithm only for real
$\lambda_i$, as I am mainly interested in the unstable mode, which is
real.


\section{Numerical error estimates}
\label{appF}

\begin{figure}
\centerline{\epsffile{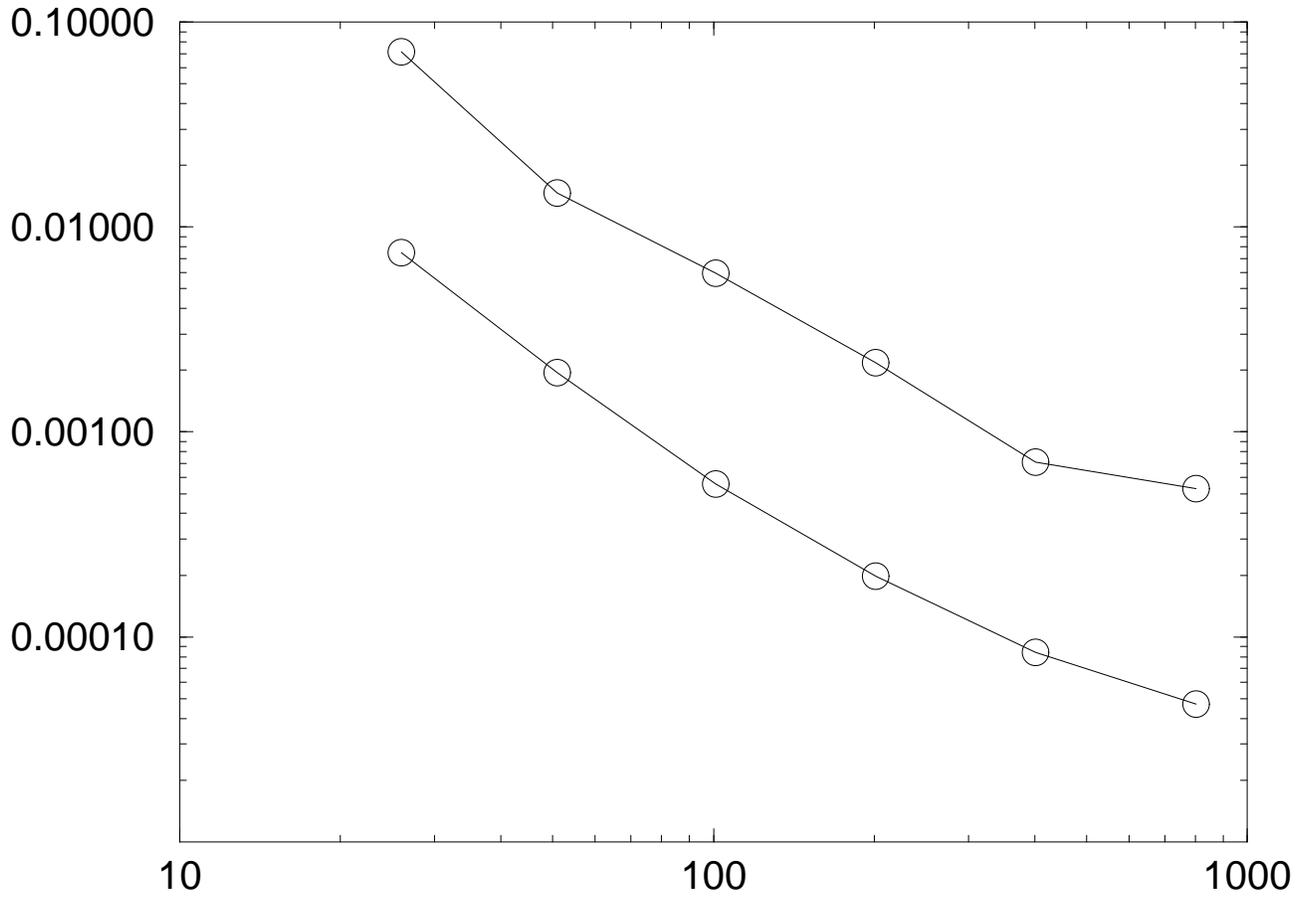}}
\caption{Convergence of the solution from $r=0$ to the past SSH. Horizontal
axis gives number of grid points (26, 51, ..., 801). Vertical axis gives
difference of numerical solution from reference solution with 1601
points. Upper line is maximum error over all components and grid points, lower
line is root-mean-square error. Upper line coincides with the error in
$\Delta$, as this is the component which has the largest error. Note that
the diagonal of the box has slope 2, the expected convergence behavior.}
\label{convergence}
\end{figure}

For an ODE discretisation to second order such as (\ref{discretized}) one
would expect the solution to converge globally to second order in the step
size. I had incorrectly claimed to observe this in \cite{PRL}. The real
situation is more complicated: the root-mean-squared or maximum measure of
convergence show convergence that is quadratic {\it on the average} over a
wide range of step sizes. By quadratic on the average I mean that the
log-log plot of step size versus the difference of two numerical solutions
with different step sizes is a somewhat wiggly line, but with an average
slope of two, as illustrated in Fig.\ \ref{convergence}. This result is
puzzling, because one would expect a much tighter result to hold, namely
that the numerical solution of the equations at a given step size $h$ is
equal to the continuum solution plus a discretisation error which itself is
a power series in even powers of $h$ only (the Richardson form). This
clearly does not hold here. I believe that the irregularity is due either
to the presence of boundary conditions, or in particular to the singular
nature of the boundaries.

The fact that the numerical solution really obeys (\ref{discretized}) to
machine precision is easily established, independently of the (complicated)
relaxation algorithm that found the solution in the first place.
Furthermore the convergence behavior should be independent of what is
inside the routines that supply the algebraic boundary conditions and the
derivative $F(f,\zeta)$. Therefore it appears unlikely that the cause of
the irregular convergence behavior is a programming error, even if such an
error was present.

When a numerical error has the Richardson form, one obtains a sharp error
estimate, and which moreover is based on a tested theory of the numerical
solution process. Here I cannot use this theory.  Nevertheless, there
clearly is something like quadratic convergence with decreasing step size,
and so I feel justified to use the difference between the numerical
solution with $N$ grid-points and the numerical solution with $2N$
grid-points as a measure of the numerical error of the solution with $N$
gridpoints. Adding a safety factor of 4, I take this value as the
error of the $2N$ solution, and the $2N$ solution itself as the best value.

Fig.\ \ref{convergence} plots maximum and root-mean-square measures of
convergence of the entire solution. As it happens, the maximum measure is
dominated by the error in $\Delta$, and so doubles as an error plot for
$\Delta$.  As described above, I obtain the value and error bar $\Delta =
3.4453 \pm 0.0005$.

Table \ref{table2} gives the values of the parameters $\Delta$, $\lambda_1$
and $\lambda_2$ for various $N$. $\lambda_2$ is the gauge mode
corresponding to a translation of the background solution, and we should
find $\lambda_2=0$. Its absolute numerical value gives a bound on the error
of $\lambda$ of $\pm 0.009$, while from the convergence of $\lambda_1$ I
would have estimated a much smaller error of $\pm 0.0002$. I opt for the
larger error estimate, and obtain $\lambda_1 = -2.674 \pm 0.009$,
corresponding to a critical exponent $\gamma = - 1/\lambda_1 = 0.374 \pm
0.001$.


\section{Curvature at the future SSH}
\label{appG}


The Riemann tensor in spherical symmetry in general coordinates has
been calculated in \cite{PoiIsr}. Substituting the form (\ref{metric})
of the metric, it is straightforward to verify that the only
appearance of any second derivatives of $a$ or $\alpha$ with respect
to $r$ and $t$ in any component of the Riemann tensor is in the
combination
\begin{equation}
{a\over \alpha}r\left(r{a_{,t}\over \alpha}\right)_{,t}
-r\left(r{\alpha_{,r}\over a}\right)_{,r}.
\end{equation}
We now use the Einstein equations 
\begin{equation}
r{a_{,t}\over \alpha} = {1\over2} a^2\left(X_+^2 - X_-^2\right),
\qquad
r{\alpha_{,r}\over \alpha} = {1\over2} (a^2-1) + {1\over 2}
a^2\left(X_+^2 - X_-^2\right),
\end{equation}
to transform the terms in round brackets into algebraic expressions,
and we transform the outer derivatives from coordinates $(r,t)$ to
coordinates $(w,\rho)$ via
\begin{equation}
gr{\partial \over \partial t} = G {\partial \over \partial w},
\qquad
r{\partial \over \partial r} = {\partial \over \partial \rho} - \Gamma
w {\partial \over \partial w}.
\end{equation}
The only derivative in the resulting expression that potentially does
not exist is $X_{+,w}$. But it arises as 
\begin{equation}
(G+\Gamma w)(a^2 X_+^2)_{,w},
\end{equation}
and while $X_{+,w}$ blows up as $(w+1)^{\epsilon-1}$ as $w\to-1$, the
coefficient vanishes as $(w+1)$, because of the coordinate condition
$\Gamma=G$ at $w=-1$. We conclude that the curvature remains finite at
$w=-1$ although not all second derivatives of the metric are finite
there. 




\begin{table}
\caption{A comparison between the Kerr-Newman and critical collapse solutions.}
\label{table1}
\begin{tabular}{cc}
Kerr-Newman solutions & Critical collapse solutions \\ 
\tableline
length scale $M$ & scale-invariant \\
stationary & self-similar \\
horizon & naked singularity \\
$\Longrightarrow$ vacuum & $\Longrightarrow$ matter-dependent \\
quasi-normal modes $e^{i \omega_i t / M}$ & perturbations $t^{\lambda_i}$ \\
attractor & attractor of co-dimension one \\
$\Longrightarrow$ ``no hair'' & $\Longrightarrow$ universality \\
\end{tabular}
\end{table}



\begin{table}
\caption{Convergence of $\Delta$ and $\lambda$ with step size in $\zeta$.
Delta is the echoing period. $\lambda_1$ is the Lyapunov exponent of the
one growing mode. Its negative inverse is the critical exponent
$\gamma$. $\lambda_2$ is the exponent of the translation gauge mode. It
must be zero and serves as a check on the numerical error. Note that the
numerical grid (and number of steps) is the same for the background as for
the perturbations in each case.}
\label{table2}
\begin{tabular}{cccc}
Number of steps & $\Delta$ & $\lambda_1$ & $\lambda_2$ \\ 
\tableline
51 &  3.4321725669119 & -2.6858281957399 & -3.8648655101422D-02 \\
101 & 3.4513015765429 & -2.6700545097384 & 6.3598135051540D-02 \\
201 & 3.4431664827331 & -2.6741157762787 & -2.4122047436405D-02 \\
401 & 3.4446384162424 & -2.6748914760819 & -2.3339465009179D-02 \\
801 & 3.4458770431665 & -2.6738878803912 & 2.9009984141595D-02 \\
1601 & 3.4453484479734 & -2.6740958070987 & -9.4268323369794D-03 \\
\end{tabular}
\end{table}



\begin{figure}
\epsfysize=16cm
\centerline{\epsffile{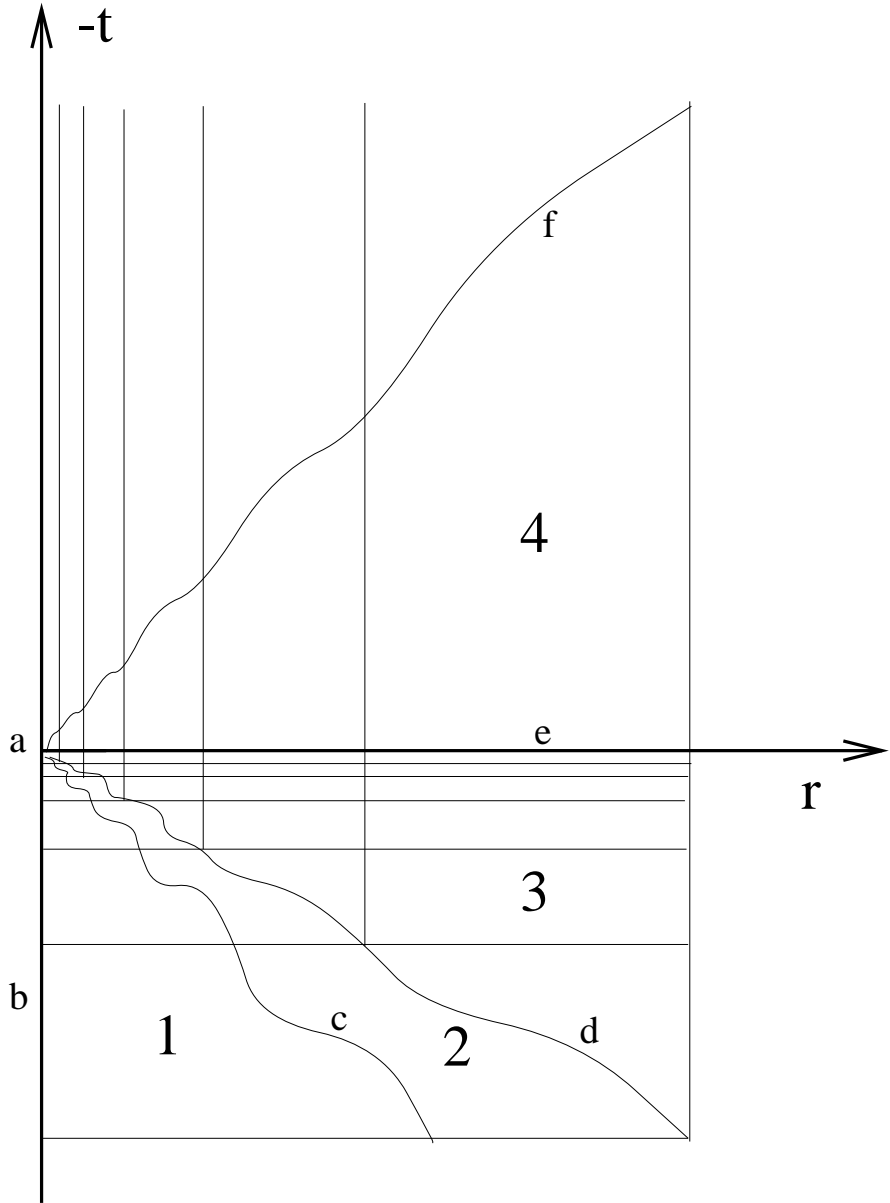}}
\caption{Schematic diagram of the coordinate systems I use. The horizontal
and vertical axes are $r$ and $-t$. Horizontal accumulating lines are
$\tau=0$, $\tau=\Delta$, $\tau=2\Delta$, etc., from bottom up. Vertical
accumulating lines are $\rho=0$, $\rho=\Delta$, $\rho=2\Delta$, etc., from
right to left.  a) The curvature singularity $(r=0,t=0)$. b) The regular
center $r=0$. c) The past self-similarity horizon $\zeta=0$. d) The
matching line $\zeta=\zeta_0$, $w=1$. e) The matching line $w=0$, $t=0$. f)
The future self-similarity horizon $w=-1$. I use four separate coordinate
patches: 1) The nonlinear eigenvalue problem for $\Delta$. 2) Continuation
to $\zeta_0$. 3) Continuation to $t=0$. 4) Continuation to the future SSH.}
\label{coordinates}
\end{figure}



\end{document}